\RequirePackage{fix-cm}
\documentclass[twocolumn,epjc3]{svjour3}
\smartqed  
\RequirePackage{graphicx}
\begin{document}

\title{Strongly screening on electron capture for Nuclides $^{52, 53, 59, 60}$Fe by the Shell-Model
Monte Carlo method in pre-supernova 
} \subtitle{}


\author{Jing-Jing, Liu\thanksref{e1,addr1},
 Qiu-He, Peng $^{2}$,
  \and Dong-Mei, Liu\thanksref{addr1} }

\thankstext{e1}{e-mail: liujingjing68@126.com}


\institute{College of Marine Science and Technology, Hainan Tropical
Ocean University, Sanya, 572022, China \label{addr1}\\
$^{2}$Department of Astronomy, Nanjing University, Nanjing, Jiangshu
210000, China \label{addr2}
}

\date{Received: date / Accepted: date}

\maketitle

\begin{abstract}
The death of the massive stars due to supernova explosion is a key
ingredient in stellar evolution, stellar population synthesis. The
electron capture (EC) plays a vital role in supernovae explosions.
According to the Shell-Model Monte Carlo (SMMC) method , basing on
the Random Phase Approximation (RPA) and Linear Response Theory
Model (LRTM), we study the strongly screening EC rates of nuclides
$^{52, 53, 59, 60}$Fe in presupernova. The results show that the
screening rates can decrease about 18.66\%. We compare our results
with those of Fuller et al. (FFN), Aufderheide et al. (AUFD), and
Nabi et al. (NKK)in the case with and without strong electron
screening(SES). For the case without SES, our calculations are in
very good agreement with those of AUFD in relatively high density
surroundings and the maximum error is within 0.35\% at $\rho_7=100$
(e. g., even-even nuclei $^{60}$Fe). However, for odd-A nuclei
$^{59}$Fe, our rates are close to one, one, two order magnitude
smaller than those of FFN, AUFD, and NKK. For the case with SES, our
screening results are about three, two orders magnitude, and 7.27\%
lower than those of FFN, AUFD, NKK for $^{60}$Fe, respectively, and
it is lower about two, two orders magnitude, and 12.42\% than those
of FFN, AUFD, NKK for odd-A nuclide $^{59}$Fe.

\keywords{stars: supernovae, stars: evolution, Physical Date and
Processes: nuclear reactions.}
\end{abstract}

\section{Introduction}

It is well known that supernovae not only plays a critical role in
the universe, but also supernovae are major sources of
nucleosynthesis of stellar evolution and galactic chemical
evolution. However, the driving mechanisms are still not well
understood for two typical supernova (i.e. the core-collapse (type
II) and thermonuclear (type Ia) supernovae). Some researches show
that electron captures(hereafter EC) on medium-heavy nuclei play an
important role in both types of supernovae. The weak interaction
(e.g., EC and Beta decay) leads up the unstable nuclear burning and
iron nucleus collapse in the process of the supernova explosions
\cite{r1,r2,r3}. Therefore, some pioneer works on EC are
investigated by Fuller, et al.\cite{r4,r5}(FFN); Aufderheide, et
al.\cite{r6,r7}(AUFD); According to the Shell-Model Monte Carlo
method\cite{r71,r72,r73,r74}, Langanke, et al.\cite{r8,r9}; and
Juodagalvis, et al. \cite{r10} in supernova. Liu, et
al.\cite{r11,r12,r13,r14,r15,r16,r17,r18,r19,r2,r3,r20,r1,r21} Nabi,
et al. \cite{r22}(NKK) also discussed the weak interaction reactions
in explosive stellar environments due to the importance of EC.

Nonetheless, the problem on supernova explosion has always been the
interesting and challenging issue in the fields of astrophysics. On
the other hand, the strong electron screening(SES) has been raised a
strong interesting and challenging problem among nuclear
astrophysicist in pre-supernova stellar evolution and
nucleosynthesis.

In the process of supernova explosion, what role on earth should the
EC play in stars? How do the temperature and the density affect on
the EC rates? what role will the SES play in stars? How does SES
affect on the EC rates?  These problems show that it is extremely
important and useful to calculate accurately the EC rates for the
research of supernova explosion and numerical-simulation. It is also
extremely necessary for us to understand, solve and calculate
accurately the SES and screening corrections in stellar interior for
the relativistic degenerate electron liquid.

Some works \cite{r6,r7,r23,r24,r14,r16,r4,r5} show that nuclides
$^{52, 53, 59, 60}$Fe are very important and dominated during the
process of supernova explosions. Thus the rates of them are widely
investigated by some eminent scholars (e.g. \cite{r6,r7,r23,r24}) in
supernova. In the same environment, Liu, et al. and Gutierrez, et
al.\cite{r18,r25} also discussed the weak interaction reactions on
$^{52, 53, 59, 60}$Fe. However, their works seem not to consider the
influence of SES on EC. The problems about SES has already been
discussed by Bravo et al.\cite{r26}, and Liu et al.\cite{r27}. The
works mentioned above show that it is extremely important and
necessary to calculate accurately the screening corrections to the
EC rates in dense stars.

The effects of the charge screening on the nuclear physics(e.g.
electron capture and beta decay) Come at least from three factors.
Firstly, the screening potential changes the electron Coulomb wave
function in the process nuclear reaction. Secondly, the electron
screening potential decreases the energy of incident electron
joining the capture reaction, which generally also decreases the
electron capture rates of the nuclei. Thirdly, the electron
screening increases the energy of atomic nucleus(i.e. the single
particle energy will increases) in the process of nuclear reaction,
thus increases the nuclear reaction rates. Finally, the electron
screening evidently and effectively decreases the number of the
higher-energy electrons, of which the energy is more than the
threshold of the capture reaction. The screening relatively
increases the threshold of the reaction and decreases the capture
rates, but increases for the beta decay rates.

Due to the importance of SES about the EC of $^{52, 53, 59, 60}$Fe
in astrophysical environments, in this paper, we focus on these
nuclei and reinvestigate their EC rates according to the Shell-Model
Monte Carlo (hereafter SMMC) method, and the theory of Random Phase
Approximation (hereafter RPA) \cite{r23}. For the case
 without SES, we analyse the error factor $C$ and compared our
 results($\lambda_{ec}^{0}$(LJ)) with those of AUFD ($\lambda_{ec}^{0}$(AUFD)),
which is based on the theory of ¡±Brink Hypothesis¡± \cite{r6,r7}.
Furthermore, we discuss the electron capture cross section
(hereafter ECCS) and the rates of the change of electron fraction
(hereafter RCEF) in process of EC by using the theory of RPA. The
comparisons of our rates with those of FFN \cite{r8}, AUFD
\cite{r6},and NKK \cite{r22} are presented in table format in the
case without SES. On the other hand, basing on the linear response
theory \cite{r28}, we investigate the strongly screening rates and
the screening factors $C_{1}$. In order to understand the influence
of SES on EC, we also compare our screening rates with those of FFN,
AUFD, and NKK. We find the influence of SES on the rates is very
significant.

Our work differs from previous works \cite{r6,r7,r4,r5,r22} about
the discussion of EC. Firstly, the works of FFN and AUFD are based
on the theory of Brink Hypothesis(BH). Basing on quasi-particle
random phase approximation, NKK also discussed the EC reaction in
the case without SES. We analyze the EC process for iron group
nuclei and derive new screening results by using the Shell-Model
Monte Carlo (SMMC) method \cite{r23} and basing on LRTM \cite{r28}
and RPA theory\cite{r8,r9}. We also make detailed comparison of the
results for the strong screening rates and no-screening rates among
the calculations by FFN, AUFD, and NKK. Secondly, our discussions
differs from Ref.\cite{r29}, which analyzed the EC by using the
method of BH and basing on the plasma ion ball strong screening
model(PIBSSM). PIBSSM is a very rough model and BH also is a very
poor approximation, which assumes that the Gamow-Teller strength
distribution on excited states is the same as for the ground state,
only shifted by the excitation energy of the state. Finally, we
analyze the effect of SES by the linear response theory model(LRTM)
These screening rates of iron group nuclide may be universal, very
important and helpful for the researches of supernova explosion and
numerical simulation.

The present paper is organized as follows: in the next section, we
analyses the EC rates in stellar interiors in the case with and
without strong electron screening. Some numerical results and
discussion are given in Section 3. And some conclusions are
summarized in Section 4.

\section[]{The EC in stellar interiors}

\subsection{The EC process in the case without SES}
The stellar electron capture rates for the $k$ th nucleus (Z, A) in
thermal equilibrium at temperature $T$ is given by a sum over the
initial parent states $i$ and the final daughter states $f$
\cite{r4,r5,r6,r7}
\begin{equation}
\label{eq.1}
  \lambda_{k}^0=\lambda_{ec}^0=\sum_{i}\frac{2(J_i+1)\exp(\frac{-E_i}{kT})}{G(Z,A,T)} \sum_{f}\lambda_{if}
\end{equation}
The EC rate from one of the initial states to all possible final
states is $\lambda_{if}$. The $J_i$ and $E_i$ are the spin and
excitation energies of the parent states,  $G(Z,A,T)$ is the nuclear
partition function and given by
\begin{equation}
\label{eq.2} G(Z,A,T)=\sum_i(2J_i+1)\exp(-\frac{E_i}{kT})
\end{equation}

Using the level density formula, $\vartheta(E, J, \pi)$ , the
contribution from the excite states is discussed. Thus the nuclear
partition function approximately becomes \cite{r7}
\begin{eqnarray}
\label{eq.3}
G(Z,A,T) &\approx& (2J_0+1)+ \int_0^\infty dE \int _{J,\pi}dJd\pi (2J_i+1)\nonumber\\
&&\times \vartheta(E, J, \pi)\exp(-\frac{E_i}{kT})
\end{eqnarray}
where the level density  is given by \cite{r12,r34}
\begin{equation}
\label{eq.4} \vartheta(E, J,
\pi)=\frac{1}{\sqrt{2\pi}\psi}\frac{\sqrt{\pi}}{12a^{\frac{1}{4}}}\times\frac{\exp[2\sqrt{a(E-\delta)}]}{(E-\delta)^{\frac{5}{4}}}
f(E, J, \pi)
\end{equation}
where
\begin{equation}
f(E, J,
\pi)=\frac{1}{2}\frac{(2J+1)}{2\psi^2}\exp[-\frac{J(J+1)}{2\psi^2}]
\label{eq.5}
\end{equation}
where $a$ is the level density parameter, $\delta$ is the backshift
(pairing correction).  $\psi$ is defined as
\begin{equation}
\psi=(\frac{2m_uAR^2}{2\hbar^2})^{\frac{1}{2}}[\frac{(E-\delta)}{a}]^{\frac{1}{4}}
\label{eq.6}
\end{equation} where $R$ is the radius and
$m_u=\frac{1}{N_A}$ is the atomic mass unit.

Based on the RPA theory with a global parameterization of the single
particle numbers, the EC rates in the case without SES is related to
the electron capture cross-section  by \cite{r10}
\begin{equation}
\lambda_{ec}^0(\rm{LJ})=\frac{1}{\pi^2\hbar^3}\sum_{if}\int^{\infty}_{\varepsilon_0}
p^2_e\sigma_{ec}(\varepsilon_e,\varepsilon_i,\varepsilon_f)f(\varepsilon_e,U_F,T)d\varepsilon_e
\label{eq.7}
\end{equation}
where $\varepsilon_0=\max(Q_{if}, 1)$. $p_e=\sqrt{\varepsilon_e-1}$
is the momenta of the incoming electron, and
 $\varepsilon_e$ is the total rest mass and kinetic energies of the
incoming electron, $U_{F}$ is the electron chemical potential, $T$
is the electron temperature. Note that in this paper all of the
energies and the momenta are respectively in units of $m_e c^2$ and
$m_e c$, where $m_e$ is the electron mass and $c$ is the light speed
in vacuum. The electron Fermi-Dirac distribution is defined as
\begin{equation}
  f=f(\varepsilon_{e},U_F,T)=[1+\exp(\frac{\varepsilon_{e}-U_F}{kT})]^{-1}
\label{eq.8}
\end{equation}

Due to the energy conservation, the electron, proton and neutron
energies are related to the neutrino energy, and $\rm{Q}$-value for
the capture reaction\cite{r32}
\begin{equation}
  Q_{i,f}=\varepsilon_{e}-\varepsilon_{\nu}=\varepsilon_{n}-\varepsilon_{\nu}=\varepsilon^{n}_{f}-\varepsilon^{p}_{i}
\label{eq.9}
\end{equation}
and we have
\begin{equation}
  \varepsilon^{n}_{f}-\varepsilon^{p}_{i}=\varepsilon^{\ast}_{if}+\hat{\mu}+\Delta_{np}
\label{eq.10}
\end{equation}
where $\hat{\mu}=\mu_{n}-\mu_p$, the difference between neutron and
proton chemical potentials in the nucleus and
$\Delta_{np}=M_{n}c^2-M_{p}c^2=1.293Mev$, the neutron and the proton
mass difference. $Q_{00}=M_{f}c^2-M_{i}c^2=\hat{\mu}+\Delta_{np}$,
with $M_{i}$ and $M_{f}$ being the masses of the parent nucleus and
the daughter nucleus respectively; $\varepsilon^{\ast}_{if}$
corresponds to the excitation energies in the daughter nucleus at
the states of the zero temperature.

The electron chemical potential is found by inverting the expression
for the lepton number density

\begin{equation}
  n_e=\frac{8\pi}{(2\pi)^3}\int^\infty_0 p^2_e(f_{-e}-f_{+e})dp_e
\label{eq.11}
\end{equation}

where $f_{-e}=[1+\exp(\frac{\varepsilon_{e}-U_{F}}{kT})]^{-1}$ and
$f_{+e}=[1+\exp(\frac{\varepsilon_{e}+U_{F}}{kT})]^{-1}$ are the
electron and positron  distribution functions respectively, $k$ is
the Boltzmann constant.

According to the Shell-Model Monte Carlo method, which discussed the
Gamow-Teller strength distributions, the total cross section by EC
is given by \cite{r23}
\begin{eqnarray}
\sigma_{ec}&=& \sigma_{ec}(E_e)=\sum_{if}\frac{(2J_{i}+1)\exp(-\beta E_i)}{Z_A}\sigma_{fi}(E_e)\nonumber\\
&=& 6g^{2}_{wk}\int d\xi(E_{e}-\xi)^2 \frac{G^2_A}{12\pi}
S_{GT^+}(\xi) F(Z,E_e)\nonumber\\
\label{eq.12}
\end{eqnarray}
where $\beta=1/T_N$ is the inverse temperature, $T_N$ is the nuclear
temperature and in unit of Mev, and $E_e=\varepsilon_e$ is the
electron energy. $S_{GT^+}$ is the Gamow-teller(GT) strength
distribution, which is as a function of the transition energy $\xi$.
The $g_{wk}=1.1661\times 10^{-5}\rm{Gev^{-2}}$ is the weak coupling
constant and $G_A$ is the axial vector form-factor which at zero
momentum is $G_A=1.25$. $F(Z, \varepsilon_e)$ is the Coulomb wave
correction which is the ratio of the square of the electron wave
function distorted by the coulomb scattering potential to the square
of wave function of the free electron.

The SMMC method is also used to calculate the response function
$R_A(\tau)$ of an operator $\hat{A}$ at an imaginary-time $\tau$. By
using a spectral distribution of initial and final states
$|i\rangle$ and $|f\rangle$ with energies $E_i$ and $E_f$.
$R_A(\tau)$ is given by \cite{r8}
\begin{small}
\begin{equation}
R_A(\tau)=\frac{\sum_{if}(2J_i+1)\exp(-\beta E_i)\exp(-\tau
(E_f-E_i))|\langle f|\hat{A}|i\rangle|^2}{\sum_i (2J_i+1)\exp(-\beta
E_i)} \label{eq:13}
\end{equation}
\end{small}
Note that the total strength for the operator is given by
$R(\tau=0)$. The strength distribution is given by
\begin{small}
\begin{equation}
S_{GT^+}(E) = \frac{\sum_{if}\delta (E-E_f+E_i)(2J_i+1)\exp(-\beta
E_i)|\langle f|\hat{A}|i\rangle|^2}{\sum_i (2J_i+1)\exp(-\beta E_i)}
 \label{eq:14}
\end{equation}
\end{small}
which is related to $R_A(\tau)$ by a Laplace Transform,
$R_A(\tau)=\int_{-\infty}^{\infty}S_{GT^+}(E)\exp(-\tau E)dE$. Note
that here $E$ is the energy transfer within the parent nucleus, and
that the strength distribution $S_{GT^+}(E)$ has units of $\rm
{Mev^{-1}}$.

The presupernova EC rates in the case without SES is given by
folding the total cross section with the flux of a degenerate
relativistic electron gas \cite{r8}
\begin{eqnarray}
\lambda_{ec}^0(\rm{LJ}) =\frac{\ln2}{6163}\int^{\infty}_{0}d\xi S_{GT}\frac{c^3}{(m_{e}c^2)^5}\nonumber\\
 \int^{\infty}_{p_0}dp_{e}p^2_e(-\xi+\varepsilon_e)^2
F(Z,\varepsilon_e)f(\varepsilon_e, U_F, T)~~~(\rm{s}^{-1})
 \label{eq.15}
\end{eqnarray}
where the $\xi$ is the transition energy of the nucleus, and
$f(\varepsilon_n, U_F, T)$ is the electron distribution function.
The $p_0$ is defined as
\begin{equation}
p_0=\left\lbrace \begin{array}{ll}~\sqrt{\bigtriangleup Q_{if}^2-m_e^2c^4}~~~~~~(Q_{if}<-m_ec^2)\\
                                  ~0 ~~~~~~(\rm{otherwise}).
                             \end{array} \right.
\label{eq.16}
\end{equation}

In the case without SES, we define the error factors $C$, which
compare our results of $\lambda^{0}_{ec}$(LJ), which discussed by
method of SMMC with those of $\lambda^{0}_{ec}$(AFUD), which
calculated basing on the method of ¡±Brink Hypothesis¡± by AUFD.
\begin{equation}
C=\frac{(\lambda^{0}_{ec}\rm{(LJ)}-\lambda^{0}_{ec}\rm{(AUFD)})}{\lambda^{0}_{ec}\rm{(LJ)}}
\label{eq.17}
\end{equation}

On the other hand, the RCEF plays a key role in stellar evolution
and presupernova outburst. In order to understand how would the SES
effect on RCEF, the RCEF due to EC reaction on the $k$ th nucleus in
SES is defined as
\begin{equation}
\dot{Y^{ec}_{e}}(k)=-\frac{X_k}{A_k}\lambda_k \label{eq.18}
\end{equation}
where $\lambda_k$ is the EC rates; $X_k$ is the mass fraction of the
$k$ th nucleus and $A_k$ is the mass number of the $k$ th nucleus.

\subsection{The EC process in the case with SES}

Using the linear response theory, Itoh et al.\cite{r28} calculated
the screening potential for relativistic degenerate electrons. We
name this the linear response theory model (hereafter LRTM) with
SES. The electron is strongly degenerate in our considerable regime
of the density-temperature. The condition is expressed as
\begin{small}
\begin{equation}
 T\ll T_F=5.930\times10^9\{[1+1.018(\frac{Z}{A})^{2/3}(10\rho_7)^{2/3}]^{1/2}-1\},
 \label{eq.19}
\end{equation}
\end{small}
where $\rho_7$ is the density in units of $10^7\rm{g/cm^3}$,
$T_{\rm{F}}$ is the electron Fermi temperature, Z and A are the
atomic number and mass number of nucleus considered, respectively.

Based on the relativistic random-phase approximation, the static
longitudinal dielectric function due to the relativistically
degenerate electron liquid calculated by Jancovici et al.
\cite{r33}. The electron potential energy, which takes into account
the strong screening by the relativistically degenerate electron
liquid, is written as
\begin{equation}
V(r)=-\frac{Ze^2(2k_{\rm{F}})}{2k_{\rm{F}}r}\frac{2}{\pi}\int_0^\infty
\frac{\rm{sin}[(2k_{\rm{F}}r)]q}{q\epsilon(q,0)}dq,
 \label{eq.20}
\end{equation}
where $\epsilon(q,0)$ is Jancovici¡¯s static longitudinal dielectric
function and $k_{\rm{F}}$ is the electron Fermi wavenumber.

Using the linear response theory, \cite{r28} calculated the
screening potential for relativistic degenerate electrons. We name
this as linear response theory model (hereafter LRTM) with SES. A
more precise screening potential in LRTM is given by
\begin{equation}
D=7.525\times10^{-3}Z(\frac{10z\rho_7}{A})^{\frac{1}{3}}J(r_s,R)
(\rm{Mev}) \label{eq.21}
\end{equation}
where $J(r_s,R)$, $r_s$ and $R$ can be found in Ref. \cite{r13}. The
formula (21) is valid for $10^{-5}\leq r_s \leq 10^{-1}, 0\leq R\leq
50$ conditions, which are usually fulfilled in the pre-supernova
environment.

If the electron is strongly screened and the screening energy is
high enough in order not to be neglected in high density plasma. Its
energy will decrease from $\varepsilon$ to
$\varepsilon^{'}=\varepsilon-D$ in the decay reaction due to
electron screening. At the same time, the screening relatively
decreases the number of high energy electrons with energies higher
than the threshold energy for electron capture. The threshold energy
increases from $\varepsilon_0$ to $\varepsilon_s=\varepsilon_0+D$.
Thus the EC rates with SES becomes
\begin{eqnarray}
\label{eq.22}
\lambda^{s}_{ec}(\rm{LJ})=\frac{\ln2}{6163}\int^{\infty}_{0}d\xi S_{GT^+}\frac{c^3}{(m_{e}c^2)^5}\nonumber\\[1mm]
\int^{\infty}_{p_0}dp_{e}p^2_e(-\xi+\varepsilon_e) F(Z,\varepsilon_e)f(\varepsilon_e,U_F,T)\nonumber\\[1mm]
=\frac{\ln2}{6163}\int^{\infty}_{0}d\xi S_{GT^+}\frac{c^3}{(m_{e}c^2)^5}\nonumber\\[1mm]
\int^{\infty}_{\varepsilon_s}d\varepsilon^{'}\varepsilon{'}(\varepsilon^{'2}-1)^{\frac{1}{2}}(-\xi+\varepsilon^{'})^2
F(Z,\varepsilon^{'})f(\varepsilon_e,U_F,T)
\end{eqnarray}

We define the screening factors $C_{1}$ in the care with and without
SES in order to understand the effect of SES on the EC process as
follows:

\begin{equation}
C_{1}=\frac{\lambda^{s}_{ec}(\rm{LJ})}{\lambda^{0}_{ec}(\rm{LJ})}
\label{eq.21}
\end{equation}

\section{The results and disscusion}
Figure 1 shows the ECCS of nuclides  $^{52, 53, 59, 60}$Fe as a
function of electron energy at temperature $T_9=9, 11$. We find with
increasing of electron energy, the ECCS increases greatly. The
higher the temperature, the faster the changes of ECCS becomes. It
is because that the higher the temperature, the larger the electron
energy and electron chemical potential are. So even more electrons
will join in the EC process due to their energy is greater than the
Q-values. Furthermore, the Gamow-Teller transition would be
dominated in this process at high temperature surroundings.

As we all know, the trigger of the electron capture requires a
minimum electron energy given by the mass splitting between parent
and daughter (i.e. $Q_{\rm{if}}$). This threshold is lowered by the
internal excitation energy at finite temperature. For even-even
parent nuclei the Gamow-Teller strength centered at daughter
excitation energies of order of 2Mev at low temperatures. Therefore,
the ECCS for these parent nuclei increase drastically within the
first couple of MeV of electron energies above threshold. But for
odd-A nuclei the Gamow-Teller distribution will peak at noticeably
higher daughter excitation energies at low temperatures. So the ECCS
are shifted to higher electron energies in comparison to even-even
parent nuclei by about 3 MeV.
\begin{figure*}
\centering
    \includegraphics[width=8cm,height=8cm]{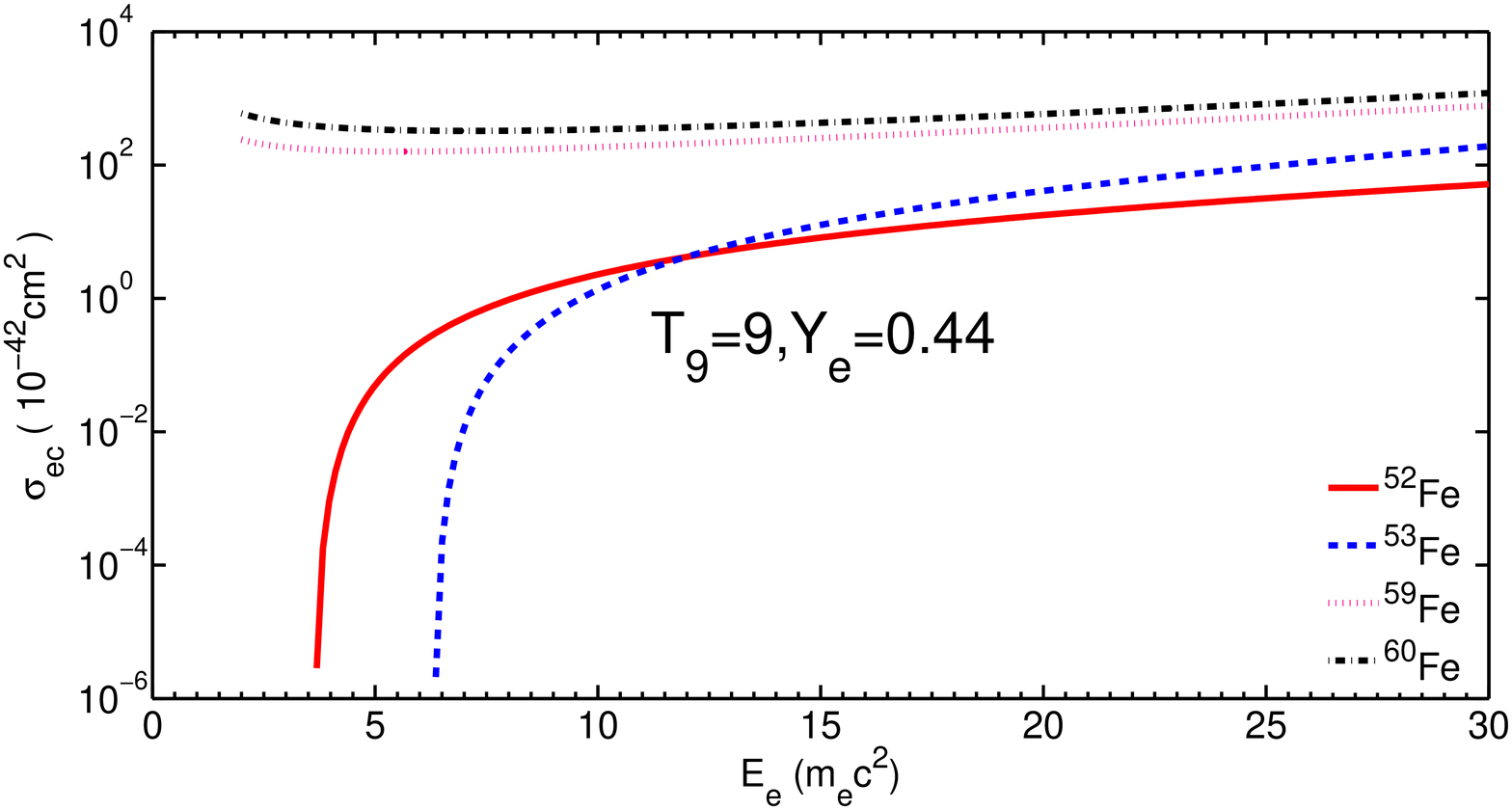}
    \includegraphics[width=8cm,height=8cm]{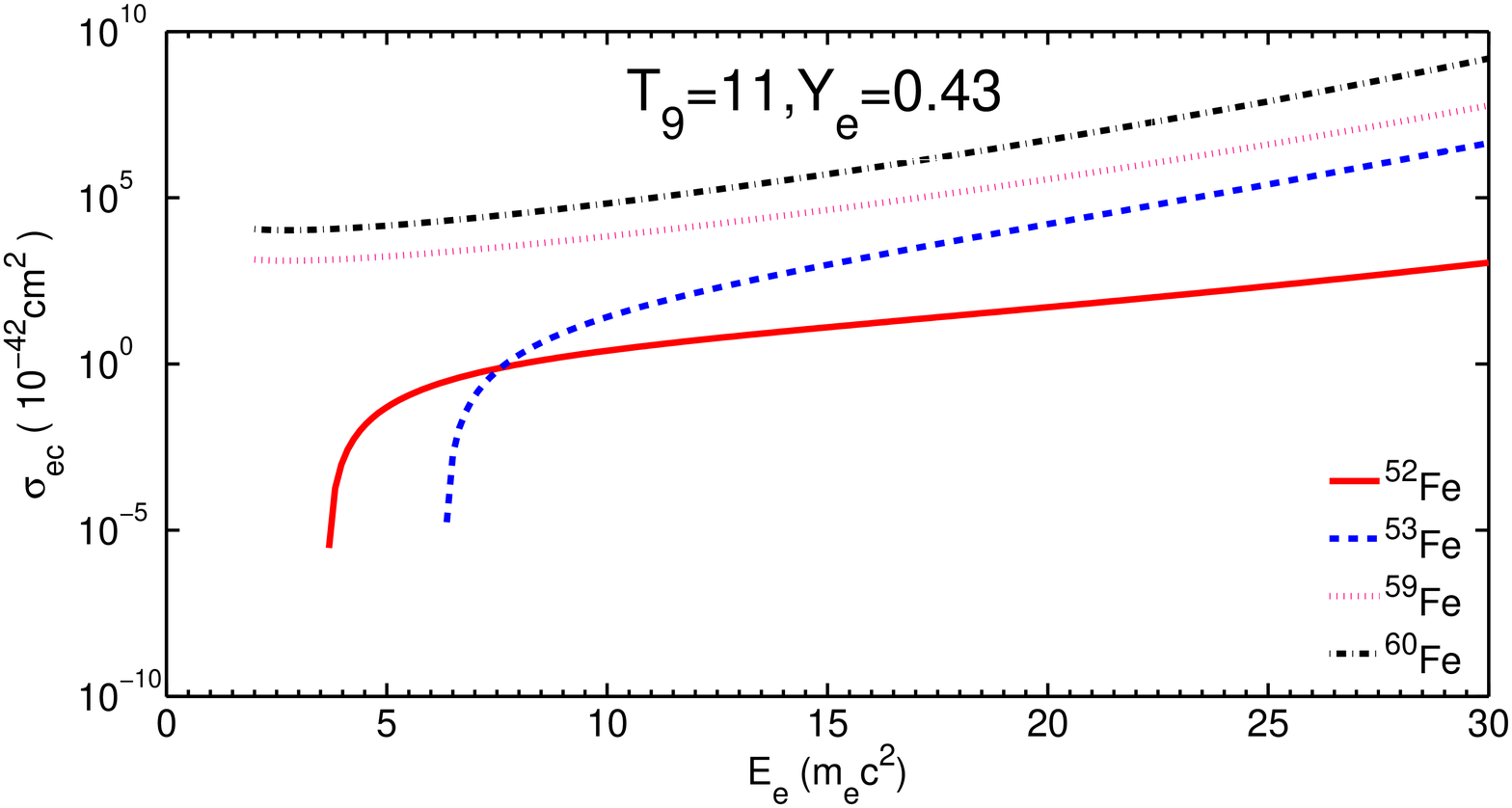}
   \caption{The ECCS for nuclides $^{52,
53, 59, 60}$Fe as a function of the electron energy at the
temperature of $T_9=9, Y_e=0.44$ and $T_9=11, Y_e=0.43$ and density
of $\rho_7=5.86$ }
   \label{Fig:1}
\end{figure*}

%
\begin{figure*}
\centering
    \includegraphics[width=8cm,height=8cm]{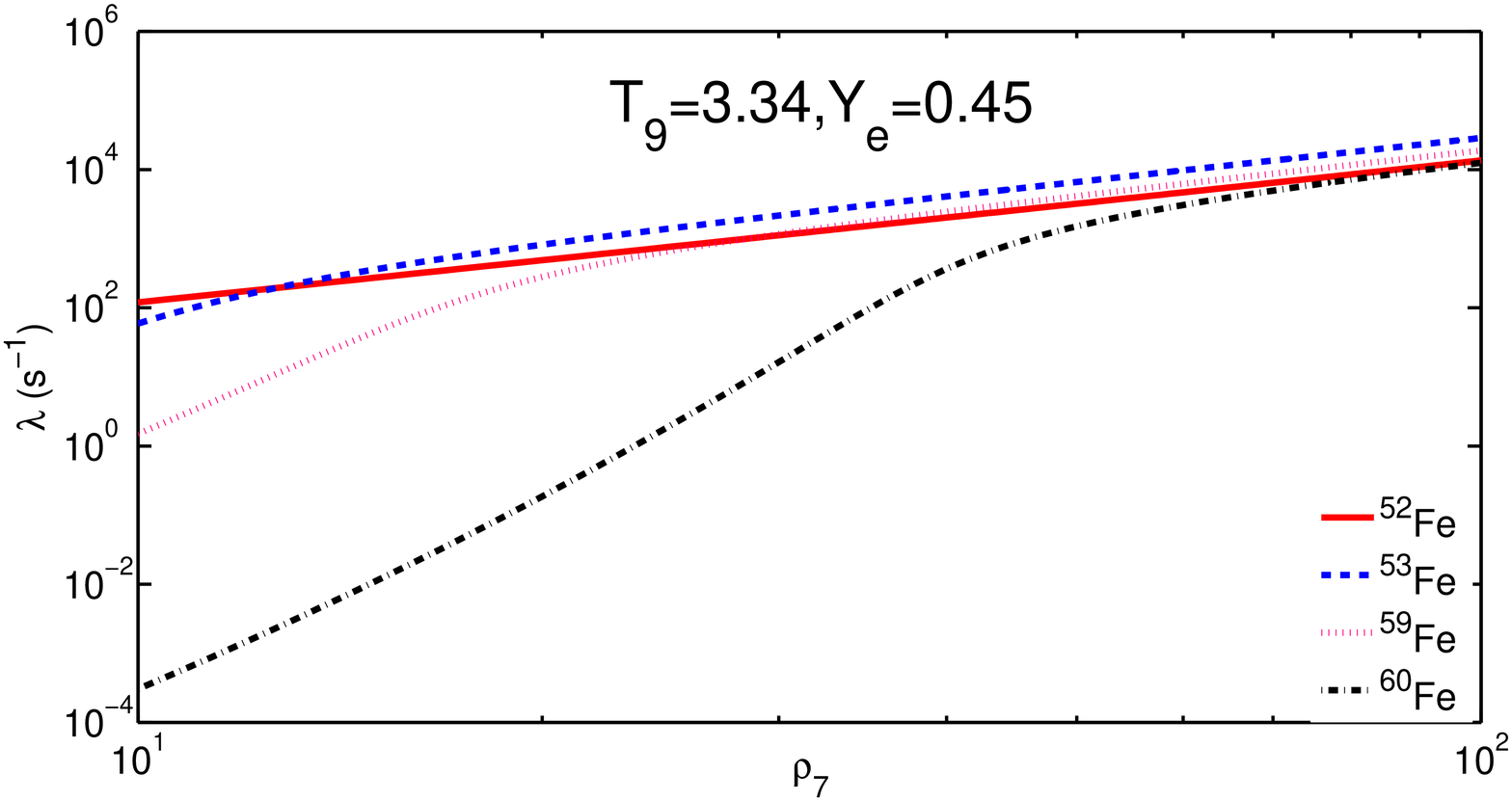}
    \includegraphics[width=8cm,height=8cm]{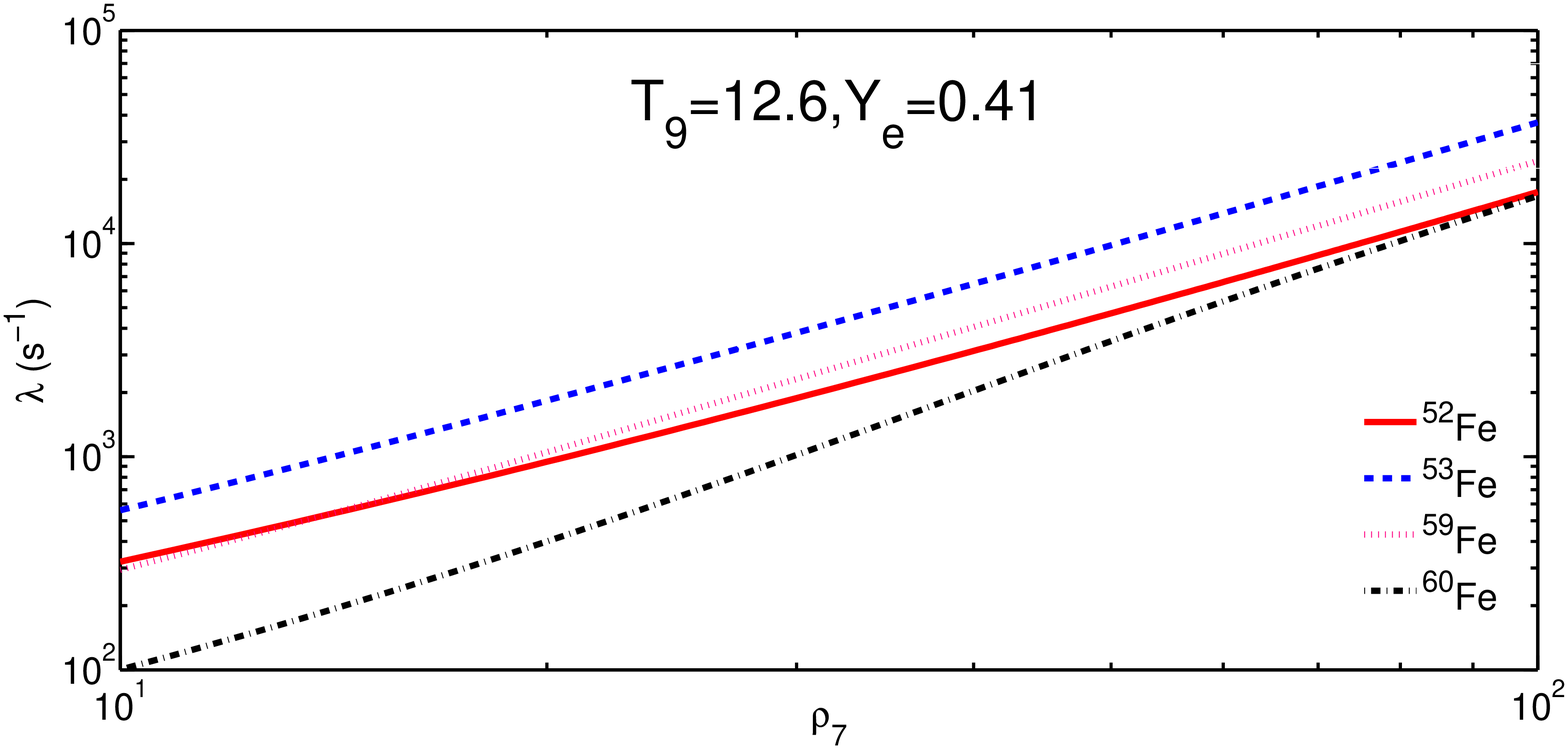}
   \caption{The EC rates for nuclides $^{52,
53, 59, 60}$Fe as a function of the density $\rho_7$ at the
temperature of $T_9=3.34, Y_e=0.45$ and $T_9=12.6, Y_e=0.41$}
   \label{Fig:2}
\end{figure*}
%
\begin{figure*}
\centering
    \includegraphics[width=8cm,height=8cm]{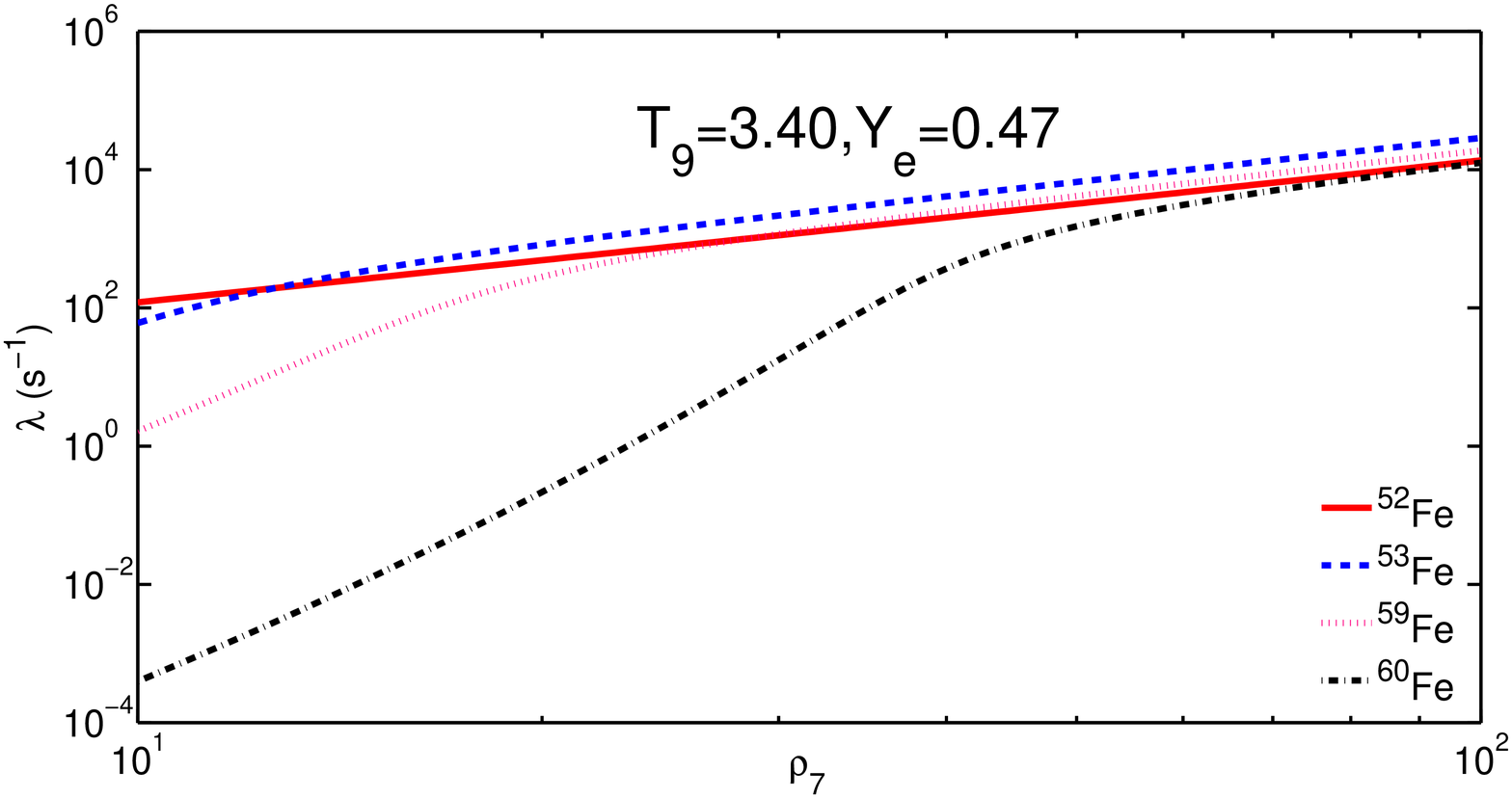}
    \includegraphics[width=8cm,height=8cm]{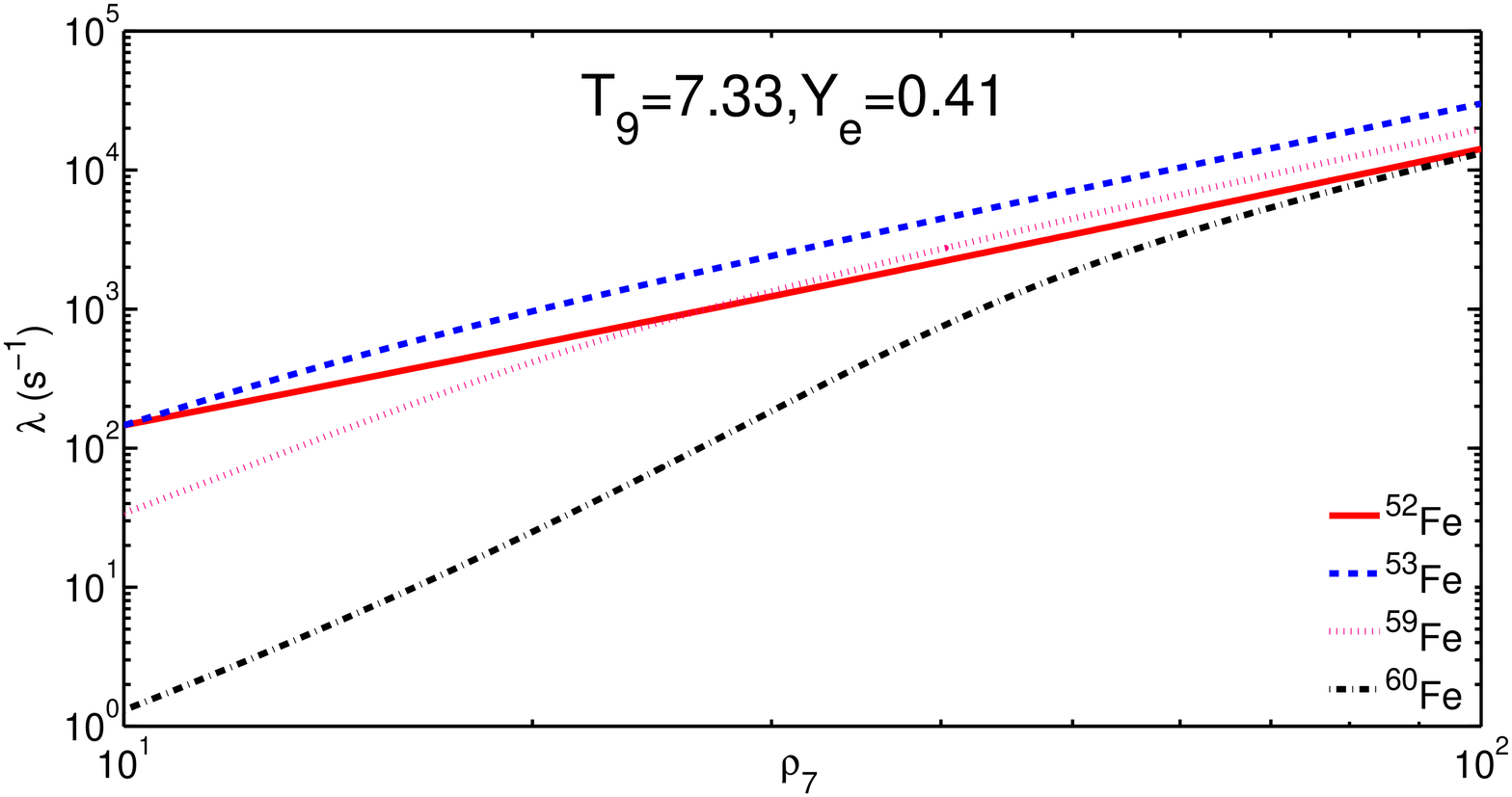}
   \caption{The EC rates for nuclides $^{52,
53, 59, 60}$Fe as a function of the density $\rho_7$ at the
temperature of $T_9=3.40, Y_e=0.47$ and $T_9=7.33, Y_e=0.41$}
   \label{Fig:3}
\end{figure*}

\begin{figure*}
\centering
    \includegraphics[width=8cm,height=8cm]{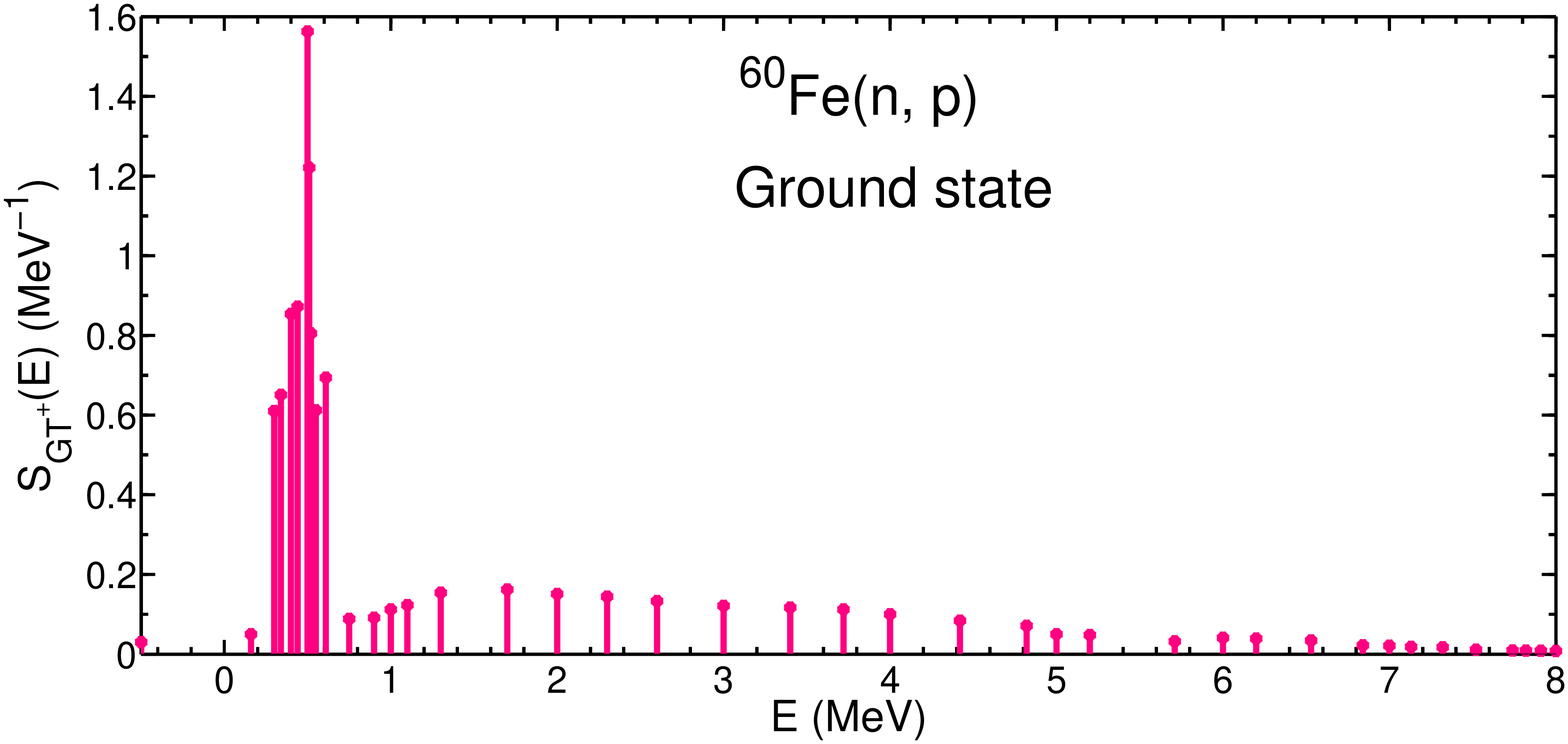}
    \includegraphics[width=8cm,height=8cm]{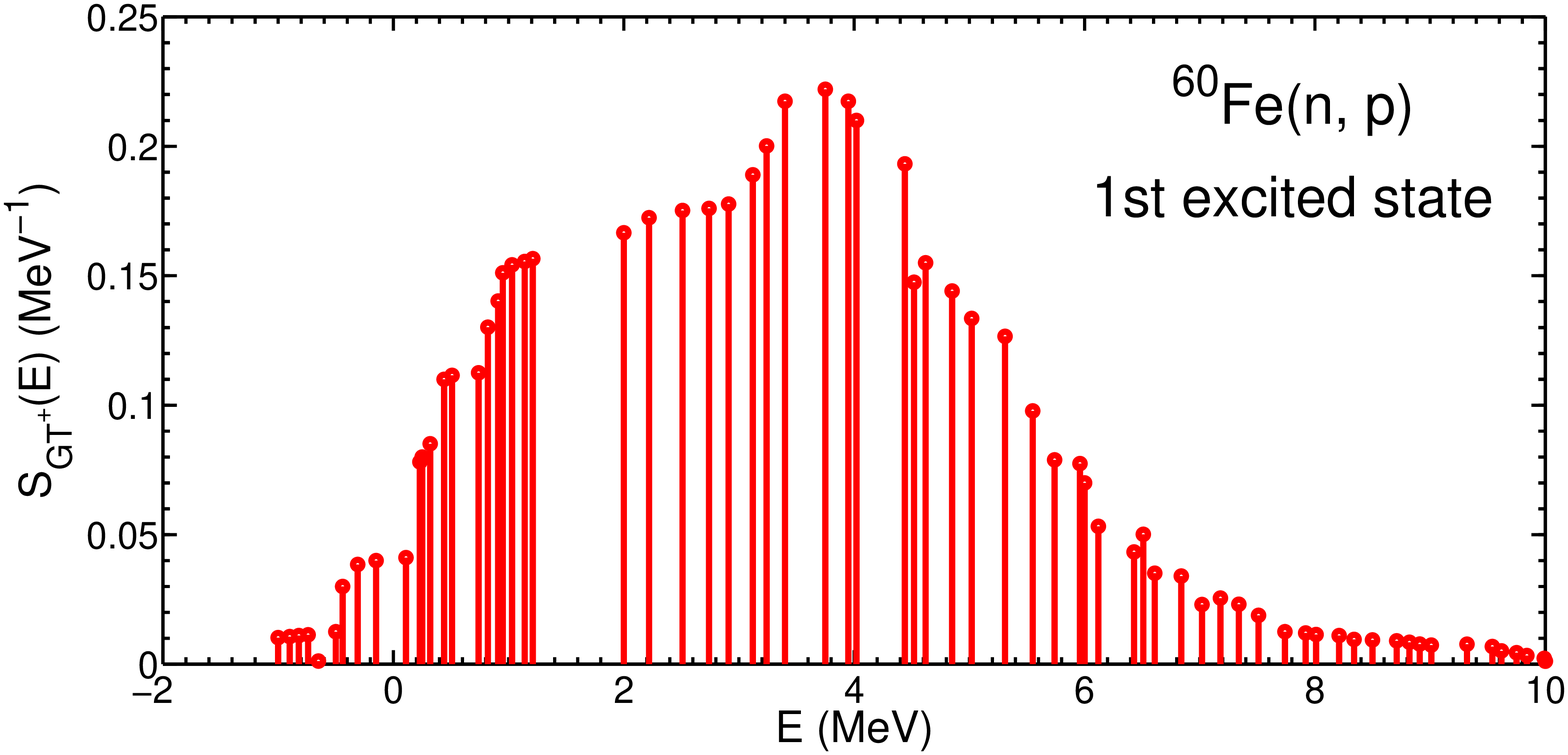}
   \caption{The theoretical $S_{\rm{GT}^+}$ for nuclei $^{60}$Fe as a function of the excitation energy $E$ at
   the ground state($0^+$) and 1st excited state($2^+$).}
   \label{Fig:3}
\end{figure*}

\begin{figure*}
\centering
    \includegraphics[width=8cm,height=8cm]{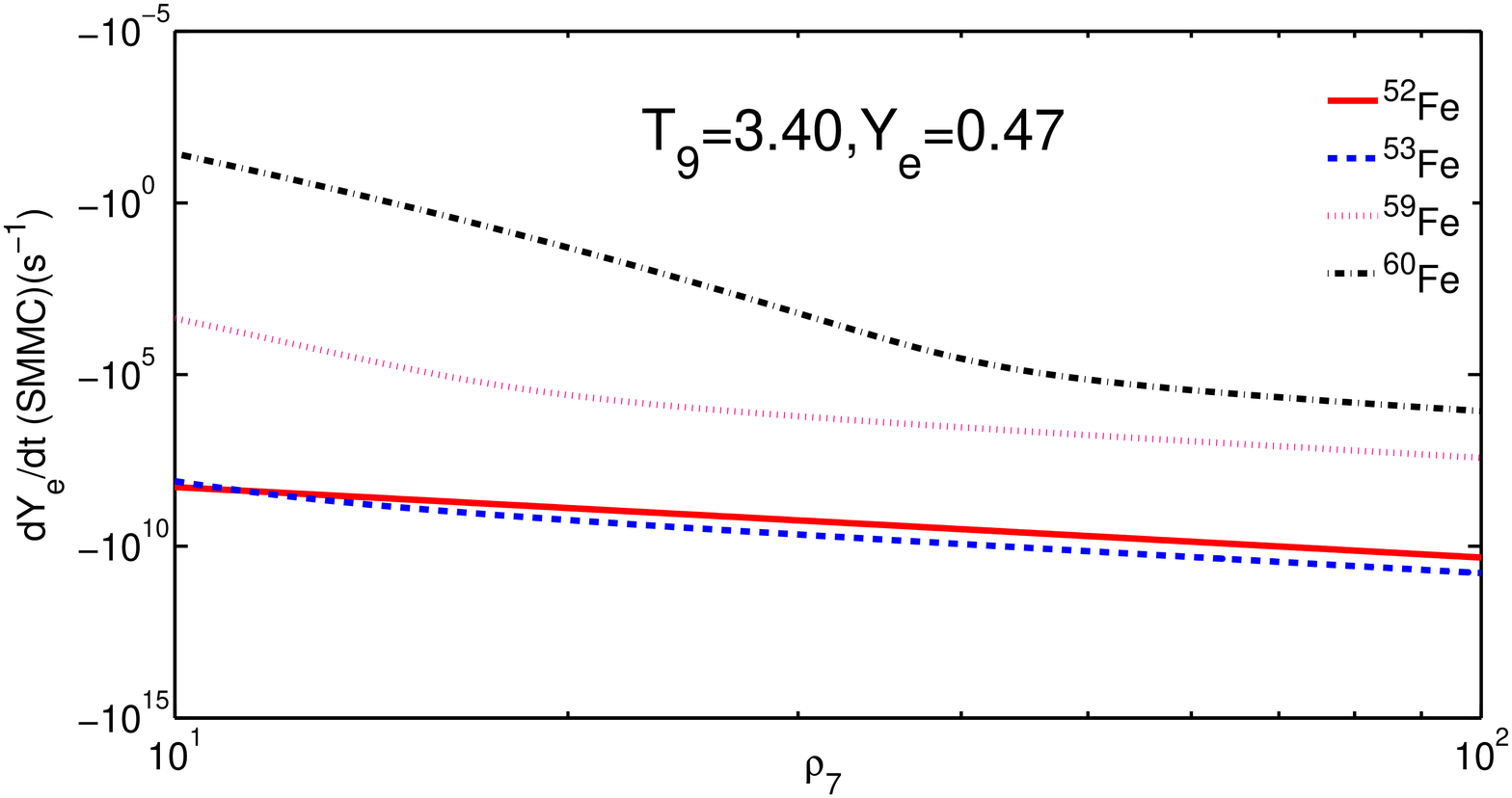}
    \includegraphics[width=8cm,height=8cm]{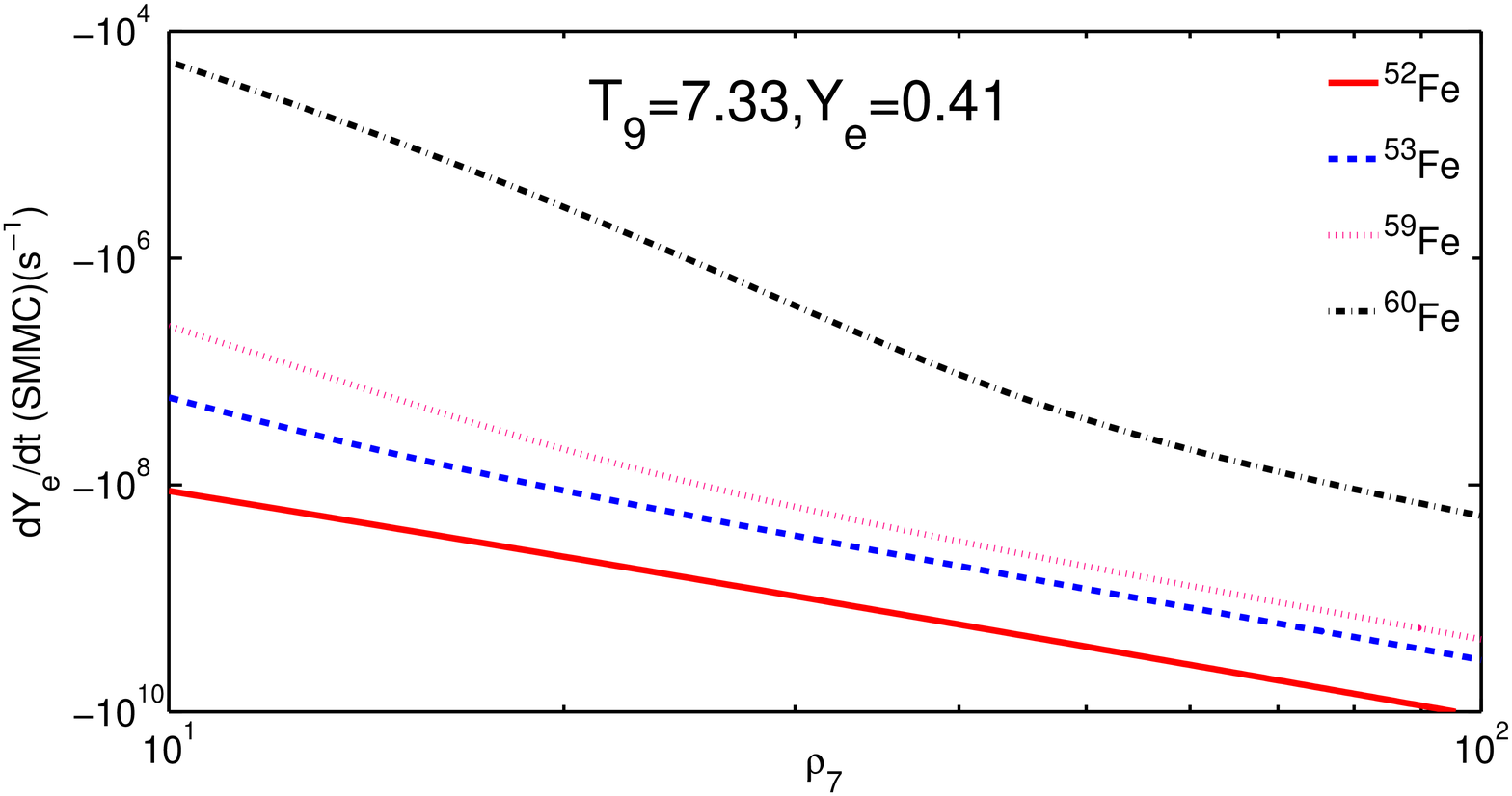}
   \caption{The RCEF due to EC process for nuclides $^{52,
53, 59, 60}$Fe as a function of the density $\rho_7$ at the
temperature of $T_9=3.40, Y_e=0.47$ and $T_9=7.33, Y_e=0.41$}
   \label{Fig:5}
\end{figure*}
\begin{figure*}
\centering
    \includegraphics[width=8cm,height=8cm]{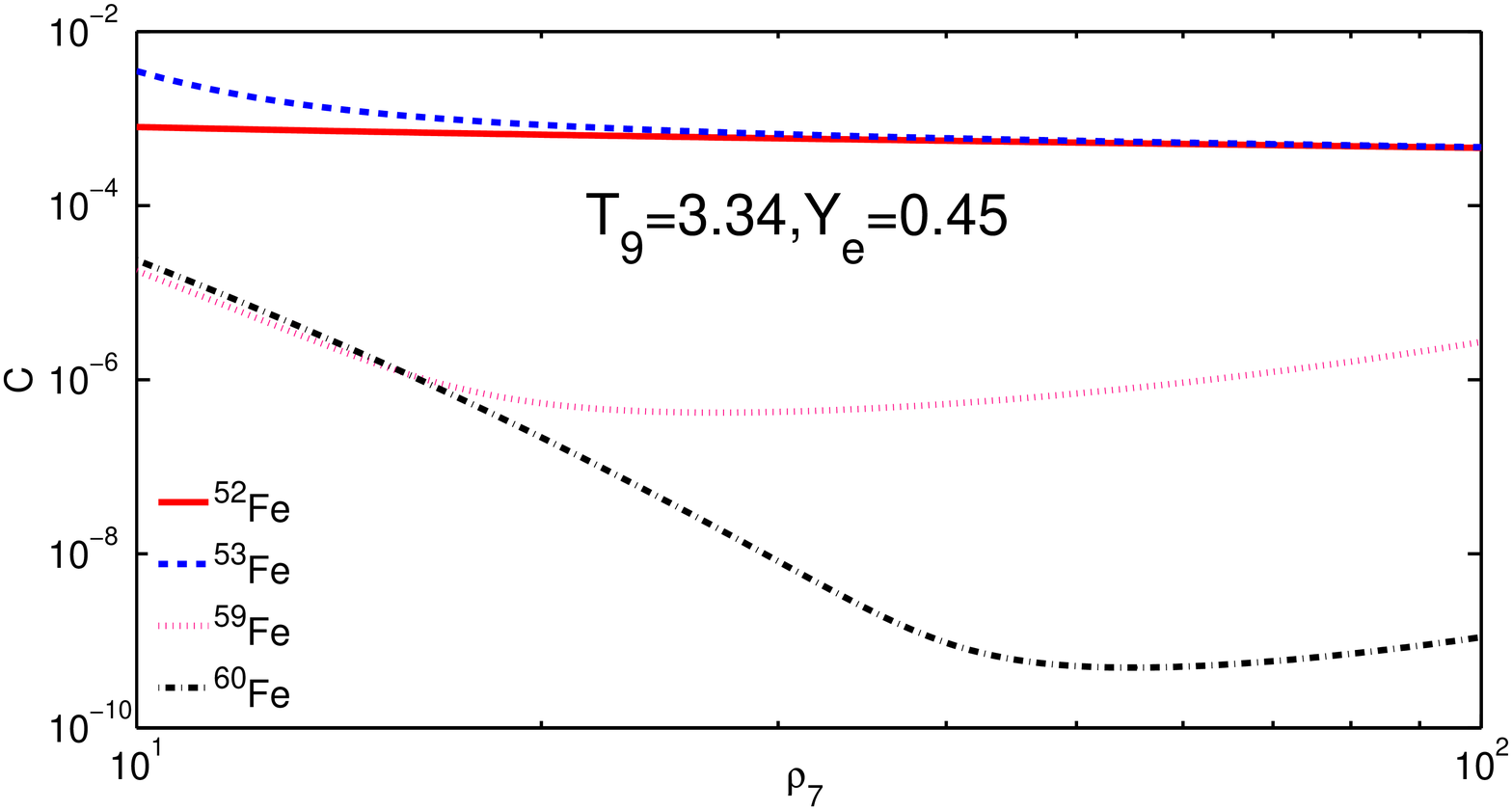}
    \includegraphics[width=8cm,height=8cm]{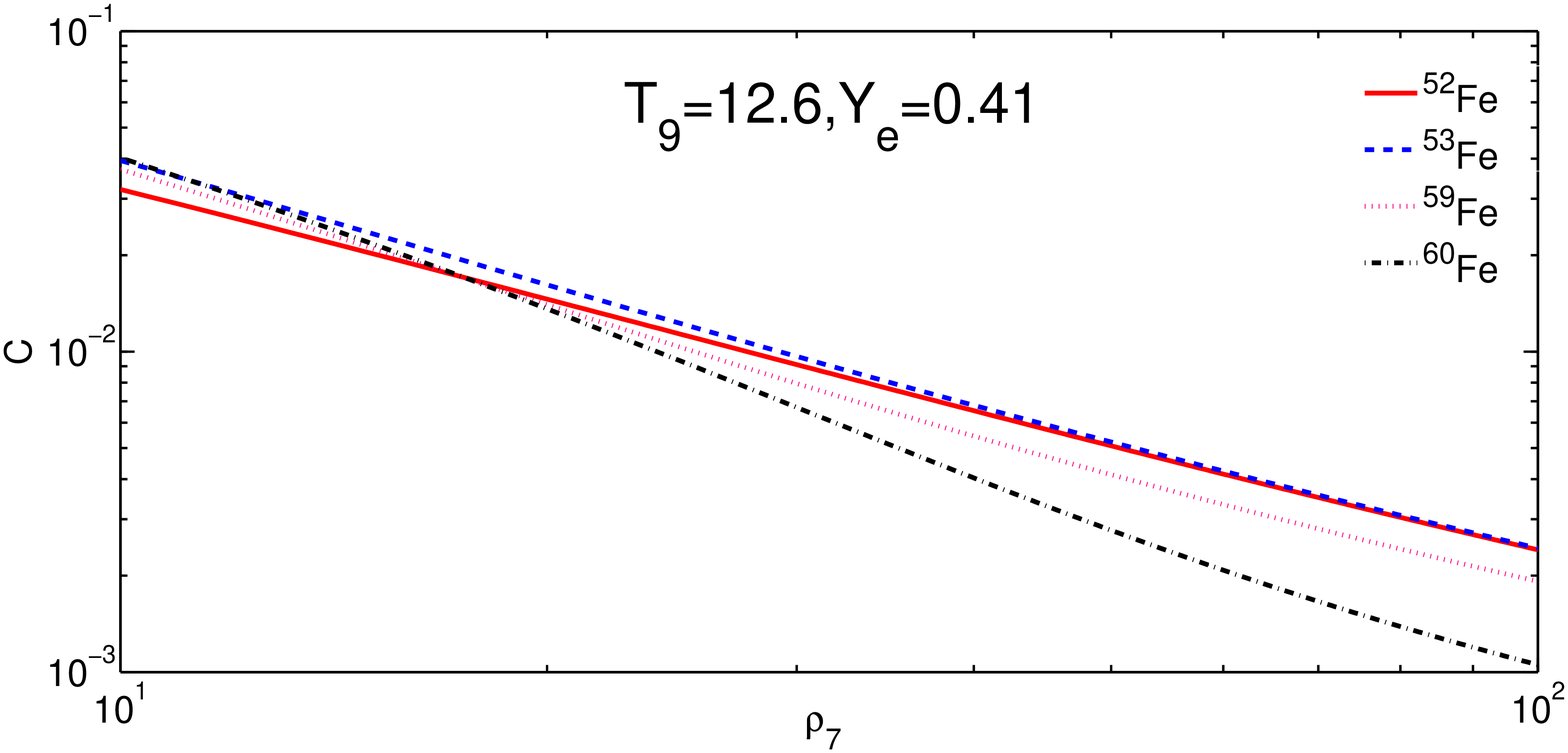}
   \caption{The factor $C$ for nuclides $^{52,
53, 59, 60}$Fe as a function of the density $\rho_7$ at the
temperature of $T_9=3.34, Y_e=0.45$ and $T_9=12.6, Y_e=0.41$}
   \label{Fig:6}
\end{figure*}

\begin{figure*}
\centering
    \includegraphics[width=8cm,height=8cm]{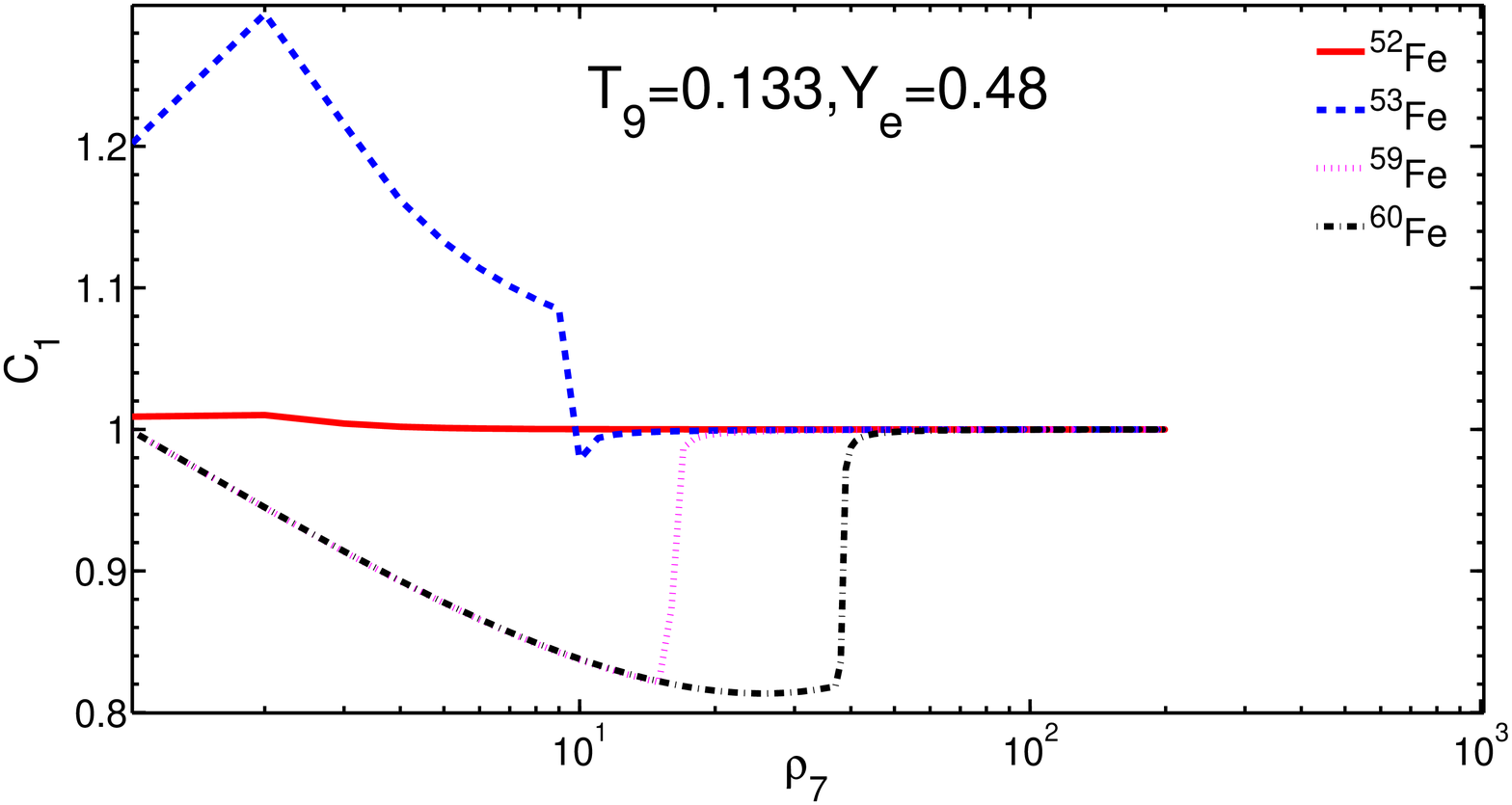}
    \includegraphics[width=8cm,height=8cm]{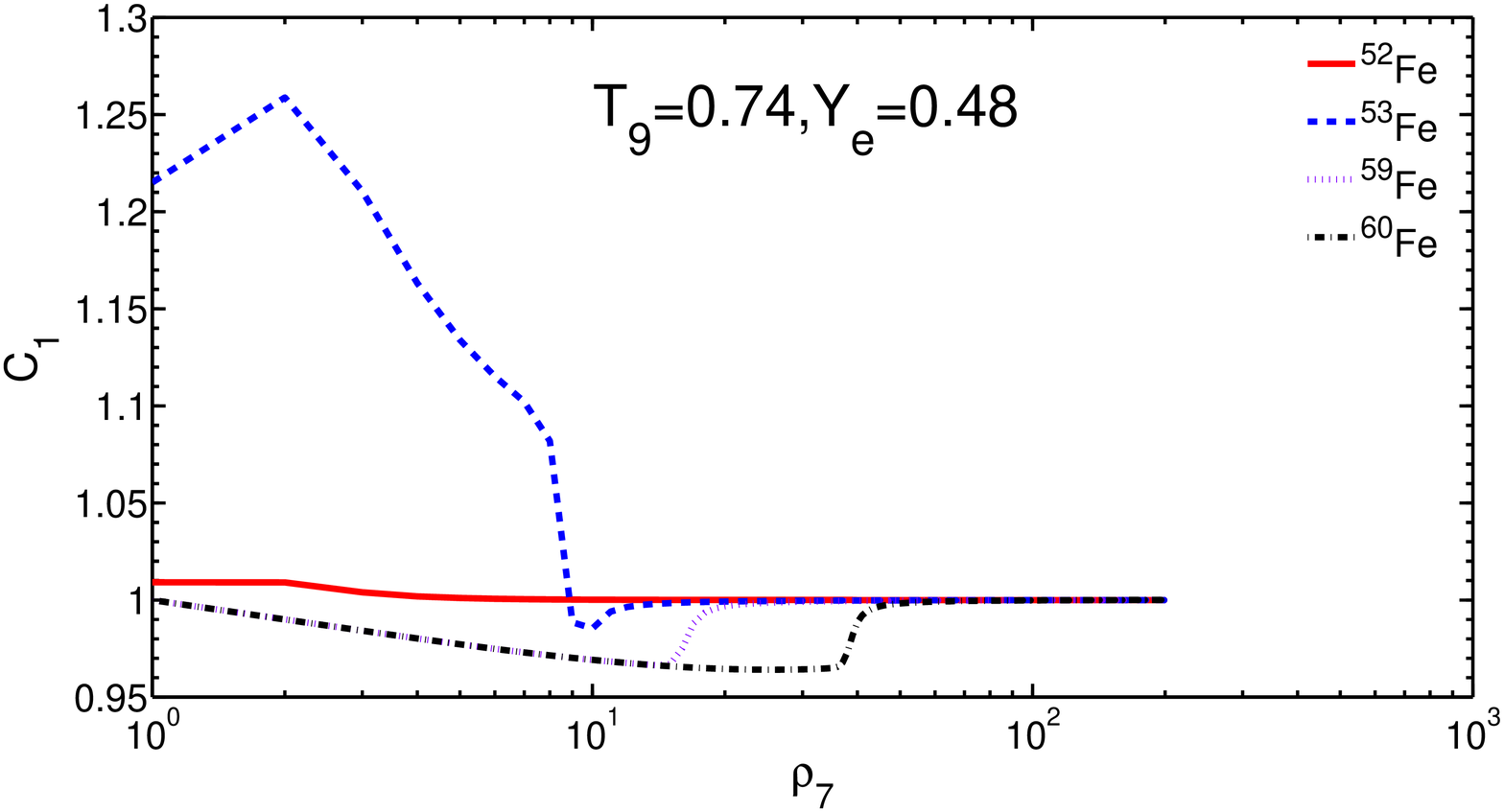}
   \caption{The screening factor $C_1$ for for nuclides $^{52,
53, 59, 60}$Fe as a function of the density $\rho_7$ at the
temperature of $T_9=0.133, Y_e=0.48$ and $T_9=0.74, Y_e=0.48$}
   \label{Fig:7}
\end{figure*}
\begin{figure*}
\centering
    \includegraphics[width=8cm,height=8cm]{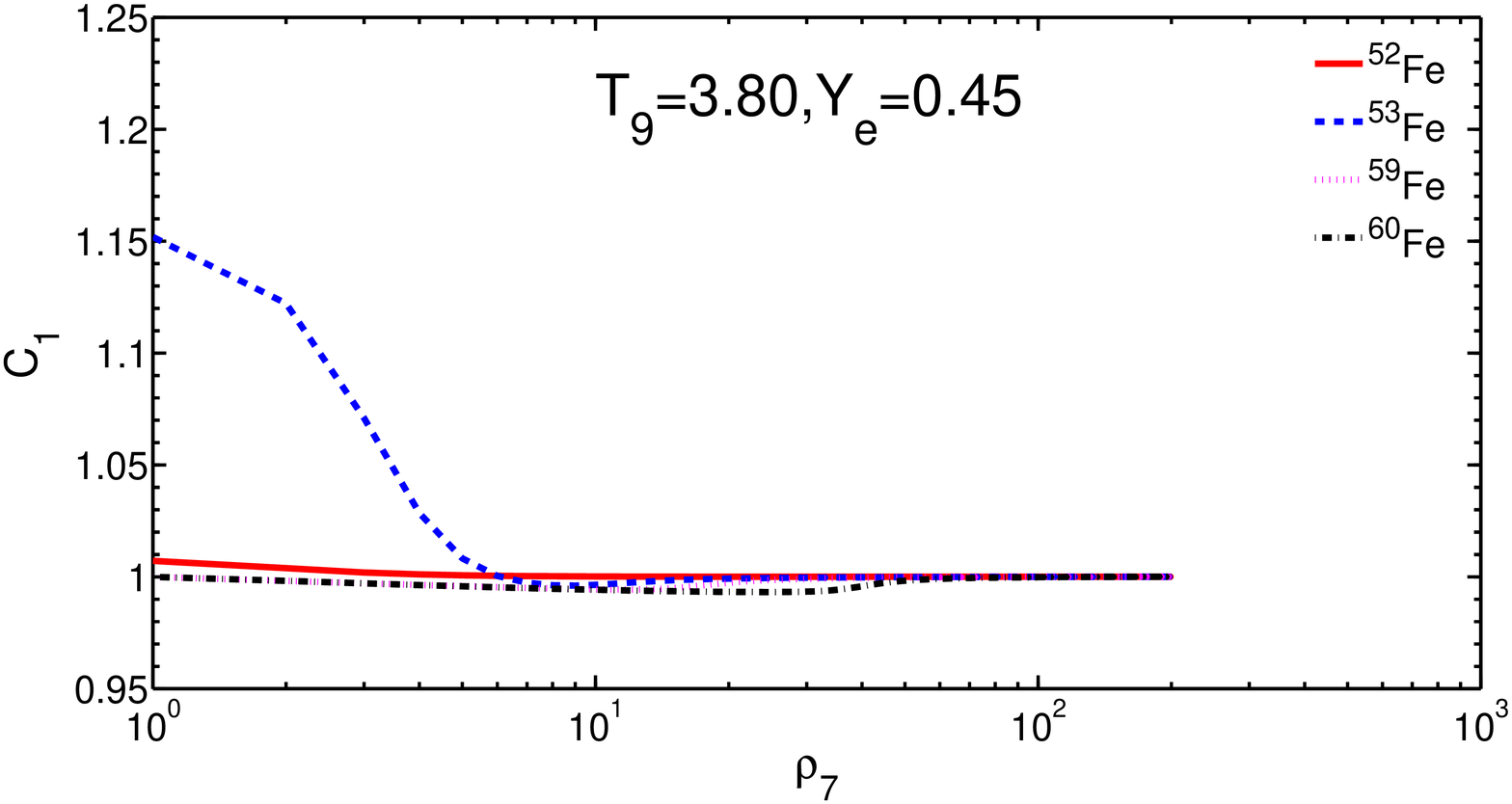}
    \includegraphics[width=8cm,height=8cm]{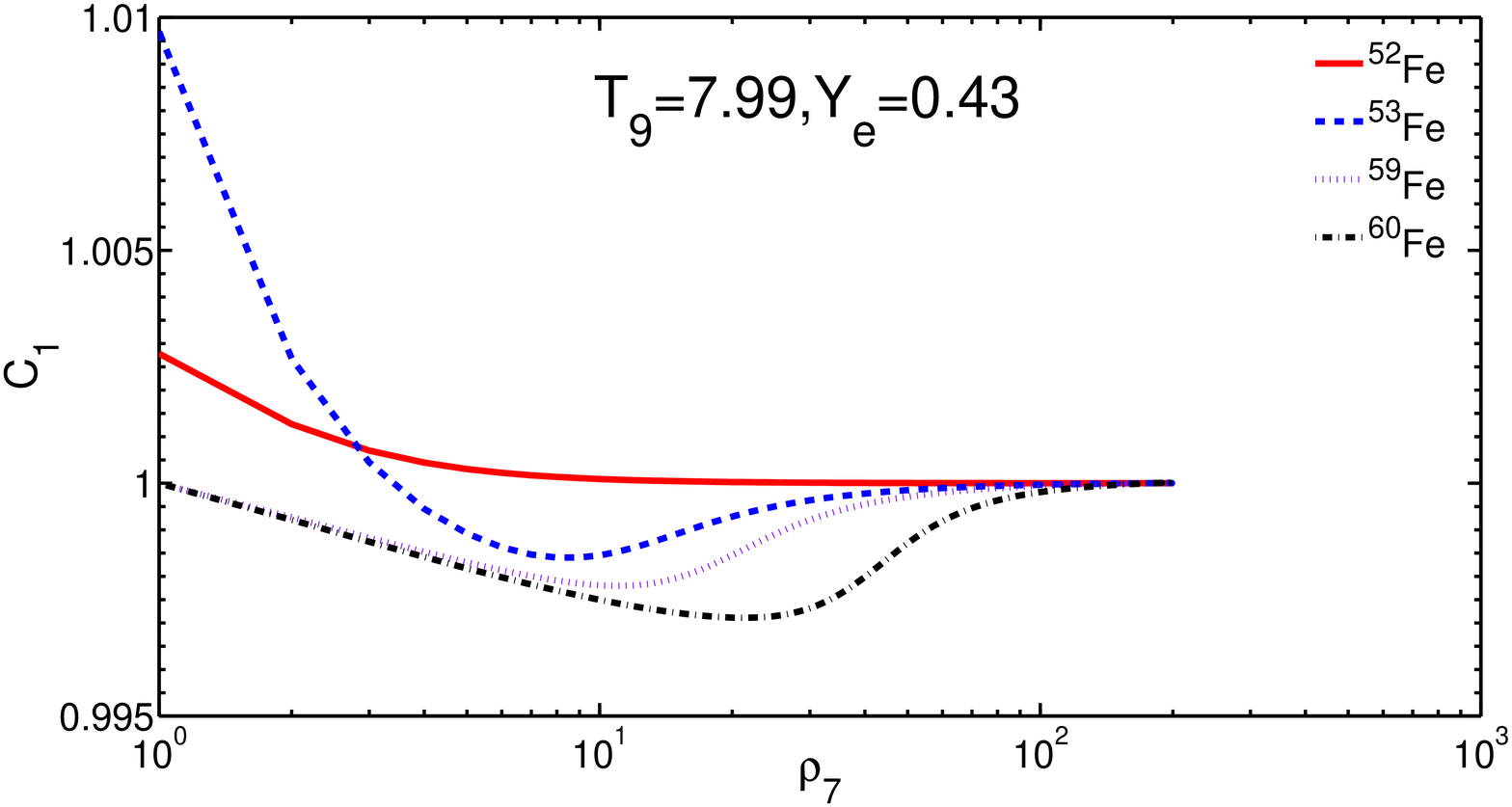}
   \caption{The screening factor $C_1$ for nuclides $^{52,
53, 59, 60}$Fe as a function of the density $\rho_7$ at the
temperature of $T_9=3.80, Y_e=0.45$ and $T_9=7.99, Y_e=0.43$}
   \label{Fig:8}
\end{figure*}

\begin{figure*}
\centering
    \includegraphics[width=8cm,height=8cm]{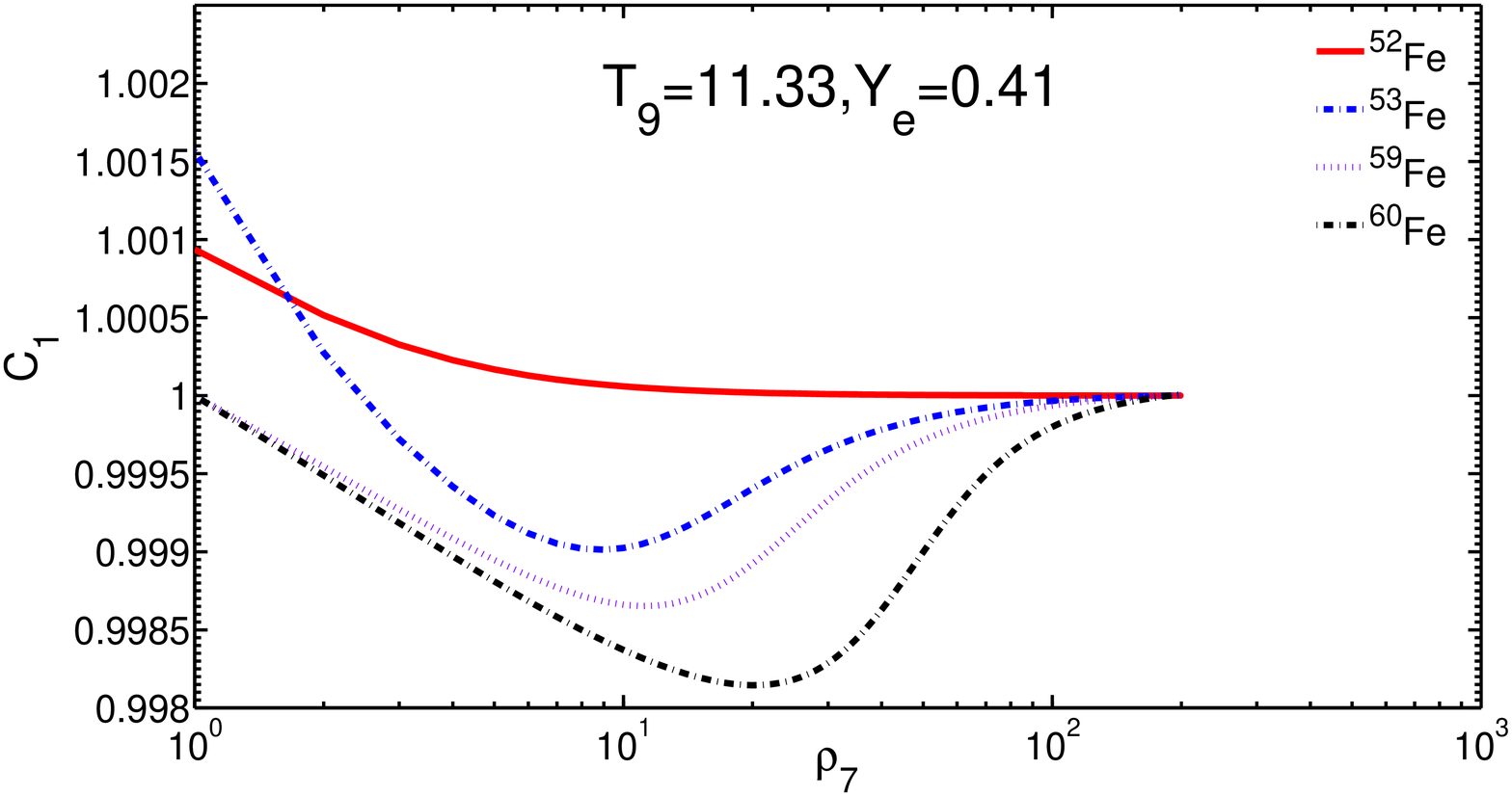}
    \includegraphics[width=8cm,height=8cm]{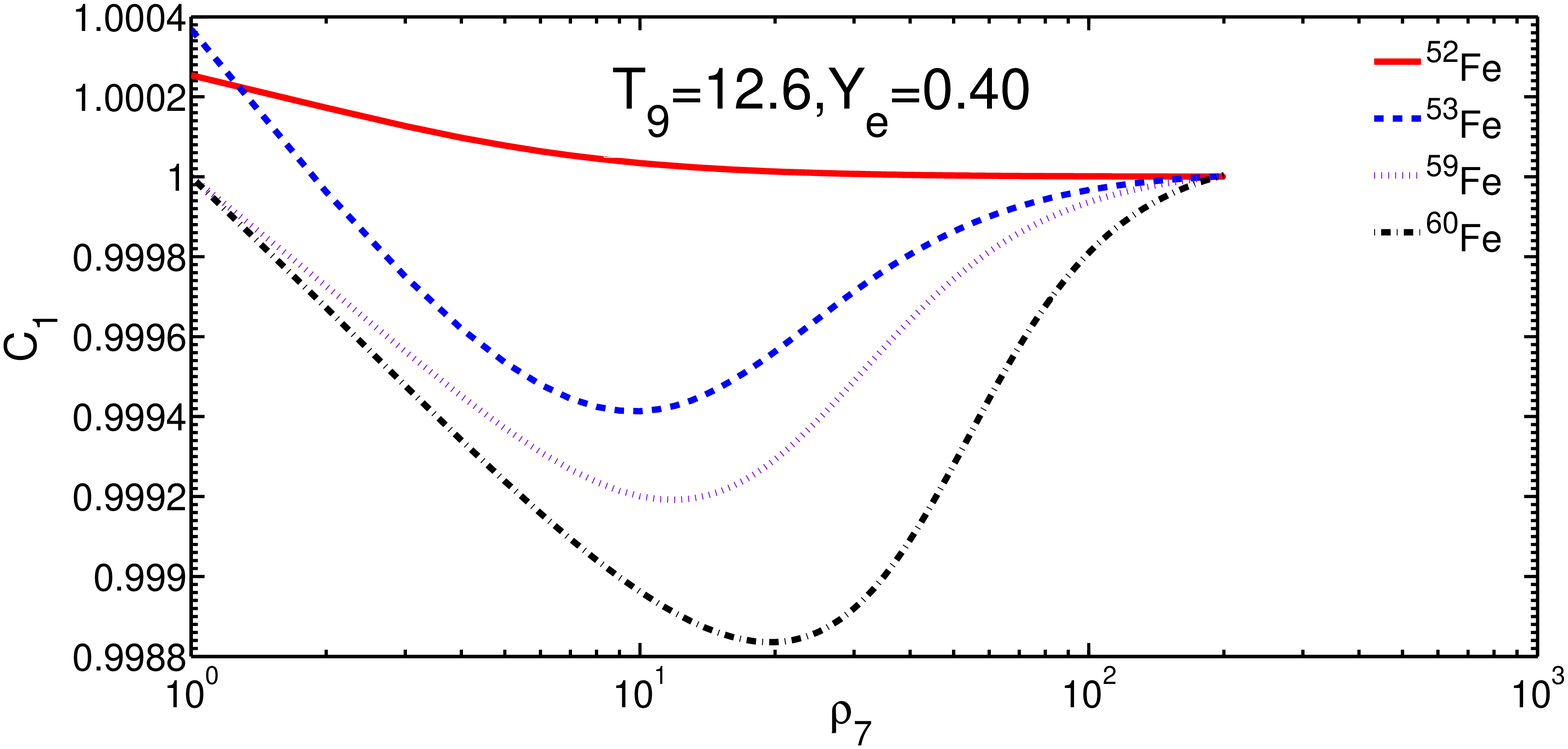}
   \caption{The screening factor $C_1$ for for nuclides $^{52,
53, 59, 60}$Fe as a function of the density $\rho_7$ at the
temperature of $T_9=11.33, Y_e=0.41$ and $T_9=12.6, Y_e=0.40$}
   \label{Fig:9}
\end{figure*}

The EC rates as function of $\rho_7$ at some typical astrophysics
condition are shown in Figure 2 and 3. We detailed discuss the EC
process according to SMMC method, especially for the contribution
for EC due to the GT transition base on RPA theory. We find the EC
rates are increased greatly and may be in excess of six orders of
magnitude with increasing of density (e. g. for $^{60}$Fe at
$T_9=3.40, Y_e=0.47$). On the other hand, the density make the
different effect on the EC rates for different nuclides due to
different Q-values and transition orbits.

According to our calculations, we find that the GT transition in EC
reaction may not be dominant at a relative low temperature. This
process can be dominated by low energy transition. On the contrary,
 the distribution of the electron gas should satisfy the Fermi-Dirac
distribution under the condition of high temperature and density.
The GT transition strength of nuclei is distributed in the form of
the centrosymmetric Gaussian function about the GT resonance point.
So the energies of the electrons, which can participate in the GT
resonance transitions  are not symmetric in relative high energy
range.

The Gamow-Teller strength distributions from shell model Monte Carlo
studies of fp-shell nuclei play an important role in the
pre-collapse evolution of supernovae.  The GT$^+$ transitions, which
change protons into neutrons, have so far been addressed only
qualitatively in presupernova simulations because of insufficient
experimental information, assuming the GT$^+$ strength to reside in
a single resonance whose energy relative to the daughter ground
state has been parametrized phenomenologically \cite{r4,r5}. (n, p)
experiments show that the GT$^+$ strength is fragmented over many
states, while the total strength is significantly quenched compared
to the single particle model.

As an example, we plot the strength distributions $S_{\rm{GT}^+}$ as
a function of daughter state excitation energy for nuclei $^{60}$Fe.
we show the calculated strength functions for GT$^+$ for the two
parent states, the ground state ($0^+$) and first excited state
($2^+$) of $^{60}$Fe in Figure 4. We consider and reproduce the
first few low-lying levels in $^{60}$Fe, which are 0, 1.1, 2.2,
2.4MeV correspondingly to the spin parity of $0^+$, $2^+$, $0^+$,
$2^+$. We find the peak of $S_{\rm{GT}^+}$ will get to 1.562,
0.223MeV$^{-1}$ at 0.5MeV, 3.40MeV of daughter nuclei $^{60}$Mn for
ground state and 1st excited state, respectively. The total GT
strength distribution $B(\rm{GT})_{tot}$ for the ground state
($0^+$) and first excited state ($2^+$) is 9.47, 8.19 MeV,
respectively. From above discussion, by simply displacing the ground
state strength distribution by the excitation energy, one can see
that the GT distribution for the excited state may be not
qualitatively inferred from the information of the ground state. In
fact we think an average value of the excited state distributions
may be the most standard distribution, which would appear to be the
one pertaining to the excited states.

The RCEF is very sensitivity parameter in the electron capture
process. We find the RCEF decreases and even more than by four
orders of magnitude for $^{60}$Fe at $T_9=7.33$ in Figures 5. With
increasing of the density and temperature, the electron chemical
potential becomes so high that large numbers of electrons join to
the EC process. Thus the RCEF reduce greatly.

It is well known that the electron capture (EC) plays a vital role
in supernovae explosions. AUFD expended  FFN's work based on shell
model. Their works are based on the theory of ¡±Brink Hypothesis¡±,
which detailed discussed by FFN in the case without SES. We analyze
the EC process and derive new results according to the Shell-Model
Monte Carlo (SMMC) method and the Random Phase Approximation (RPA)
theory. In this paper, we define the error factors C in order to
compare our results ($\lambda_{ec}^{0}$(LJ)) with those of AUFD (
$\lambda_{ec}^{0}$(AUFD)). We hope to find some difference between
these two methods at different typical stellar conditions.

Figure 6 shows the results of  the error factors C as a function of
density $\rho_7$. One can see that with increasing of density, the
factor C reduces greatly. According to our calculations, we find
that our results is agreed reasonably well with AUFD in a high
density environment (e. g. $\rho_7=100$) and the maximum error is
within 0.35\%. On the other hand, in a relative low density
surroundings the maximum error is within 3.982\% (e. g. $\rho_7=10,
Y_e=0.41, T_9=12.6$).

The EC rates on these iron group nuclei are important from the
oxygen shell burning phase up to the end of convective core silicon
burning phase of massive stars. Some pioneer works on EC rates have
been done by FFN, AUFD, and NKK. As example, the comparisons of
several EC rates(i.e. FFN's, AUFD's, NKK's, and ours) for nuclides
$^{59}$Fe and $^{60}$Fe are presented in a tabular form at
$\rho_7=4010, Y_e=0.41, T_9=7.33$ in the case without SES. We find
it is well agreement between ours and AUFD's for even-even nuclide
$^{60}$Fe. The factor C is about 0.832, 3.848, and 1.7267
corresponding to those of AUFD, FFN and NKK. The comparisons for
odd-A nuclide $^{59}$Fe show that the rates of FFN,  AUFD, and NKK
are close to by one, one, and two order magnitude bigger than ours.

Table 2 presents the comparison of our strongly screening results
with those of FFN, AUFD, NKK. From the results of $s_i$ (i=1, 2, 3,
4), one can conclude that the strongly screening rates are about
three and two orders magnitude lower than those of FFN and AUFD for
even-even nuclide $^{60}$Fe, respectively, but is about two orders
magnitude for odd-A nuclide $^{59}$Fe. On the other hand, due to
SES, our screening rates decreases about 12.42\%, 7.27\% comparing
to those of NKK for odd-A nuclide $^{59}$Fe and even-even nuclide
$^{60}$Fe, respectively.

The screening factors $C_{1}$ is plotted as a function of $\rho_7$
from figure 7 to 9. Due to SES, we find the rates decrease greatly
and even more than by $\sim 18.66$\% and $\sim 17.80$\% in Figure 6.
The lower the temperature, the larger the effect on EC rates
becomes. This is due to the fact that the SES mainly decreased the
number of higher energy electrons, which can actively join in the EC
reaction. On the other hand, the lower the temperature(e. g. in
Figure 7), the larger the effect on $C_{1}$ is. However, the higher
temperature(e. g., in Figure 8), the higher the average electron
energy becomes,  thus the smaller the effect on $C_{1}$ is, due to
the relatively low screening potential. One can also find from
Figure 7 to 9 that the screening factor is nearly the same at higher
density and independent on the temperature and density. The reason
is that at higher density surroundings the electron energy is mainly
determined by its Fermi energy, which is strongly decided by
density.

Because of relative low electron screening potential at the low
density, we find that the lower the density, the smaller the effect
on EC becomes. As the density increases, the $C_{1}$ increases
gradually due to the increases of the screening potential. As the
density further increases, the factor $C_{1}$ will close to
identical at relative high density. This is because the electron
energy is mainly determined by Fermi energy at higher density and
the effect is relatively weakened by temperature. As the density
increases, the electronic Fermi and shielding potential increases.
The ratio between shielding potential and Fermi energy has nothing
to do with density approximatively.

Of course, we know the screening of nuclear electric charges with a
high electron density means a short screening length, which means a
lower enhancement factor from Coulomb wave correction. However, even
a relatively short electric charge screening length will not have
much effect on the overall rate due to the weak interaction is
effectively a contact potential. A bigger effect is that electrons
are bound in the plasma. Table 3 in detail shows the numerical
calculations about the relationship by the minimums value of
screening factor $C_{1min}$ between the rates in the case with and
without SES. For example, the EC rates of nuclei $^{52, 53, 59,
60}$Fe are decreased about $\sim 1.40$\%, $\sim 2.12$\%, $\sim
17.80$\%, $\sim 18.66$\% at $T_9=0.133,Y_e=0.485$, respectively.

The Q-value of electron capture for some neutron rich nuclei (e.g.
$^{60}$Fe) has not been measured, so that the EC Q-value has to be
estimated with a mass formal by FFN. FFN used the Semiempirical
atomic mass formula (see Ref.\cite{r34}), thus the Q-value used in
the effective rates are quite different. On the other hand, For
odd-A nuclei (e.g. $^{59}$Fe), FFN places the centroid of the GT
strength at too low excitation energies (see the discussions in
Ref.\cite{r5}). Their rates are thus somewhat overestimated. Using
the nuclear shell model, AUFD expanded the FFN's works. AUFD
analyzed the nuclear excited level by a simple calculation on the
nuclear excitation level transitions. The capture rates are made up
of the lower energy transition rates between the ground states and
the higher energy transition rates between GT resonance states. The
works of FFN and AUFD may be an oversimplification and therefore the
accuracy is limited. they adopt the so-called Brink¡¯s hypothesis in
their calculations. This hypothesis assumes that the Gamow-Teller
strength distribution on excited states is the same as for the
ground state, only shifted by the excitation energy of the state.
This hypothesis is used because no experimental data is available
for the Gamow-Teller strength distributions from excited states and
they did not employ any microscopic theory to calculate the Gamow-
Teller strength functions from excited states.

Using the pn-QRPA theory, NKK expanded the FFN's works and analyzed
nuclear excitation energy distribution. They have taken into
consideration the particle emission processes, which constrain the
parent excitation energies. The pn-QRPA theory calculates stronger
Gamow-Teller strength distribution from these excited states
compared to those assumed using Brink¡¯s hypothesis. However in the
GT transitions considered process in their works, only low angular
momentum states are considered.

The method of SMMC is actually adopts an average of GT intensity
distribution of electron capture and the calculated results are in
good agreement with experiments, but the results for most nuclei are
generally smaller than other methods, especially for some odd-A
nuclides(e.g. $^{59}$Fe). The charge exchange reactions (p, n) and
(n, p) make it possible to observe in the process of weak
interaction, especially for the information of the total GT strength
distribution in nuclei. The experimental information is particularly
rich for some iron nuclei and it is the availability of both
$\rm{GT}^+$ and $\rm{GT}^-$, which makes it possible to study the
problem of renormalization of $\sigma\tau$ operators in detail. We
have calculated the total GT strength in a full p-f shell
calculation, resulting in
$\rm{B}(\rm{GT})=g_A^2|\langle\vec{\sigma}\tau_{+}\rangle|^2$, where
$g_{\rm{A}}^2$ is axial-vector coupling constant.

For example, in presupernova the electron capture reaction on
$^{59}$Fe is dominated by the wave functions of the parent and
daughter states. The total GT strength for $^{59}$Fe in a full p-f
shell calculation, is resulting in $\rm{B}(\rm{GT})=10.1 g_A^2$
\cite{r4}. An average of the GT strength distribution is in fact
obtained by SMMC method. A reliable replication of the GT
distribution in the nucleus is carried out and detailed analysis by
using an amplification of the electronic shell model. Thus the
method is relative accuracy.

Summing up the above discussion, basing on RPA and linear response
theory, by using the method of SMMC, we have discussed the EC rates
in SES.  One can see that the SES has an evident effect on EC rates
for different nuclei, particularly for heavier nuclides whose
threshold is negative (e.g. $^{59, 60}$Fe) at relative lower
temperature and higher density environment. According to above
calculations and discussion, one can conclude that the strongly
screening rates are decreased greatly and may be in excess of ~$\sim
18.66$\% based on the RPA theory and SMMC method.


\begin{table*}
\caption{The comparisons of our calculations of EC rates in the case
without SES for nuclides $^{59}$Fe and $^{60}$Fe with those of FFN
\cite{r5}, AUFD \cite{r7} and NKK \cite{r22} at $\rho_7=4010,
Y_e=0.41, T_9=7.33$. The ratio computes as
$k_i=\frac{\lambda_{ec}^0(i)}{\lambda_{ec}^0(\rm{LJ})}$,
$\lambda_{ec}^0(i)$ ($i=1, 2, 3$) is the rates for FFN, AUFD, and
NKK respectively in the care without SES.} \label{t.lbl}
\begin{center}
\begin{tabular}{@{}lccccccr@{}}
\hline\noalign{\smallskip}
Nuclide & $\lambda_{ec}^0$(FFN)  & $\lambda_{ec}^0$(AUFD) & $\lambda_{ec}^0$(NKK)  & $\lambda_{ec}^0(\rm{LJ})$  & $k_1$   & $k_2$ & $k_3$   \\
 \hline\noalign{\smallskip}
$^{59}$Fe  & 7.20e$+$02 & 7.43e$+$02 & 2.7e$+$02 & 7.789e$+$01    & 9.244    & 9.539& 34.704\\ 
$^{60}$Fe  & 6.73e$+$01 & 1.44e$+$01  &3.02$+$01& 1.749e$+$01    & 3.848    & 0.823&  1.7267\\
\noalign{\smallskip}\hline
\end{tabular}
\end{center}
\end{table*}
\tabcolsep 0.06in
\begin{table*}
\caption{The comparisons of our calculations of EC rates for
nuclides $^{59}$Fe and $^{60}$Fe with those of FFN\cite{r5}, AUFD
\cite{r7} and NKK \cite{r22} at $\rho_7=33, Y_e=0.45, T_9=4.24$. The
ratio computes as
$s_j=\frac{\lambda_{ec}^s(\rm{LJ})}{\lambda_{ec}^0(j)}$,
$\lambda_{ec}^0(j)$ ($j=1, 2, 3, 4$) is the rates for FFN, AUFD,
NKK, and ours respectively in the care without SES.} \label{t.lbl}
\begin{center}
\begin{tabular}{@{}lccccccccr@{}}
\hline\noalign{\smallskip}
Nuclide & $\lambda_{ec}^0$(FFN)  & $\lambda_{ec}^0$(AUFD) & $\lambda_{ec}^0$(NKK)  & $\lambda_{ec}^0(\rm{LJ})$ & $\lambda_{ec}^s(\rm{LJ})$ & $s_1$   & $s_2$& $s_3$ & $s_4$   \\
 \hline\noalign{\smallskip}
$^{59}$Fe  & 6.30e$-$03 & 5.30e$-$03  & 6.20e$-$05  & 5.63e$-$05  & 5.43e$-$05 & 8.6190e$-$03    & 1.0245e$-$02  & 0.8758  & 0.9644 \\ 
$^{60}$Fe  & 4.60e$-$03 & 1.00e$-$03  & 1.10e$-$05  & 1.08e$-$05  & 1.02e$-$05 & 2.2174e$-$03    & 1.0200e$-$02  & 0.9273  & 0.9444  \\
\noalign{\smallskip}\hline
\end{tabular}
\end{center}
\end{table*}

\begin{table*}
\caption{The minimums value of strong screening factor $C_1$, which
is comparisons of the screening rates with those of no-screening
rate for some typical astronomical condition when $1 \leq\rho_7 \leq
200$.} \centering
 \begin{minipage}{150mm}
  \begin{tabular}{@{}rrrrrrrrrrrr@{}}
  \hline
 & \multicolumn{2}{c}{$T_9=0.133, Y_e=0.485$} & &\multicolumn{2}{c}{$T_9=0.74, Y_e=0.481$}&&\multicolumn{2}{c}{$T_9=3.80,
 Y_e=0.45$}&&\multicolumn{2}{c}{$T_9=7.99, Y_e=0.43$}\\

\cline{2-3} \cline{5-6} \cline{8-9} \cline{11-12}\\
 Nuclide &$\rho_7$ & $C_{\rm{min}}$& & $\rho_7$ & $C_{\rm{min}}$ & &$\rho_7$ & $C_{\rm{min}}$& &$\rho_7$ & $C_{\rm{min}}$  \\

 \hline
 $^{52}$Fe  &25 &0.9986   & &18  &0.9997   & &19   &0.9998     & &41   &0.9999  \\
 $^{53}$Fe  &10 &0.9788   & &10  &0.9854   & &9   &0.9960     & &8   &0.9984      \\
 $^{59}$Fe  &15 &0.8220   & &15  &0.9670   & &13   &0.9944     & &12   &0.9978  \\
 $^{60}$Fe  &26 &0.8134   & &26  &0.9641   & &14   &0.9937     & &21   &0.9971    \\

\hline
\end{tabular}
\end{minipage}
\end{table*}

\section{Conclusions}
In this paper, basing on RPA theory and LRTM, by using SMMC method,
we have carried out an estimation for the EC rates of $^{52, 53, 59,
60}$Fe in the case with and without SES. Meanwhile, the ECCS and the
RCEF are discussed in SES. We also detailed compare our results with
those of FFN, AUFD, and NKK, which are in the case without SES.

Firstly, We find the influence on ECCS  is very obvious and
significant by temperature under the condition of SES. With
increasing of electron energy, the ECCS increases greatly. The RCEF
is very sensitivity parameter in the EC process and the RCEF
decreases and even more than by four orders of magnitude (e.g. for
$^{60}$Fe at $T_9=7.33$).

Secondly, for the case without SES, the EC rates increase greatly
and ever exceed by six orders of magnitude(e. g. for $^{60}$Fe at
$T_9=3.40, Y_e=0.47$). We compare our results with those of AFUD due
to different methods for calculating the EC rates. One can find our
calculations are in very good agreement with those of AUFD in
relative high density surroundings (e. g. $\rho_7=100$) and the
maximum error is within 0.35\%. However, it is within 3.982\% in a
relative low density surroundings(e. g. $ \rho_7=10, Y_e=0.41,
T_9=15.6$). On the other hand, as examples, we also discuss the
comparisons of our calculated rates with those of FFN, AUFD and NKK
of $^{59}$Fe and $^{60}$Fe. We find it is well agreement between our
results and AUFD's for even-even nuclide $^{60}$Fe (i. e. the factor
C is about 0.832, but is 3.848, 1.7267 corresponding to FFN and NKK,
respectively). The comparisons for odd-A nuclide $^{59}$Fe show that
the rates of FFN, AUFD, and NKK are close to by one, one, and two
order magnitude bigger than ours.

Finally, for the case with SES, by using SMMC method, we discuss the
strongly screening rates in supernovae explosive stellar
environments basing on RPA and linear response theory. We compare
our strongly screening results with those of FFN, AUFD, and NKK in
the case without SES. One can find that the strongly screening rates
are about three and two orders magnitude lower than those of FFN and
AUFD for even-even nuclei $^{60}$Fe, respectively, but it is lower
about two orders magnitude for odd-A nuclei $^{59}$Fe. Our screening
rates are decreased about 12.42\%, 7.27\% comparing to those of NKK
for odd-A nuclide $^{59}$Fe and even-even nuclide $^{60}$Fe,
respectively. However, according to our calculations, our strongly
screening rates ($\lambda_{ec}^s$(LJ)) are decreased greatly and
even exceed ~$\sim 18.66$\% corresponding to those of
$\lambda_{ec}^0$(LJ) in the case without SES.

As we all know, the EC by SES play an important role in the dynamics
process of the collapsing core of a massive star. It is main
parameter which leads to a supernova explosion and stellar collapse.
It also is quite relevant for simulations in the process of collapse
and explosion for massive star. The SES also strongly influence on
the cooling rate and evolutionary timescale in EC and beta decay
process. Our calculations may be helpful for study of the stellar
and galactic evolution and nucleosynthesis calculations. The results
we derived, may become a good foundation for the future
investigation of the evolution of late-type stars, the nature of
mechanism of supernova explosions and the numerical simulation of
supernovas.

\begin{acknowledgements}
This work is supported by the National Natural Science Foundation of
China under grants 11565020, and the Counterpart Foundation of Sanya
under grant 2016PT43, the Special Foundation of Science and
Technology Cooperation for Advanced Academy and Regional of Sanya
under grant 2016YD28, the Scientific Research Staring Foundation for
515 Talented Project of Hainan Tropical Ocean University under grant
RHDRC201701, and the Natural Science Foundation of Hainan province
under grant 114012.
\end{acknowledgements}

%
%



\begin{thebibliography}{}
%
%
\bibitem{r1} J. J. Liu, \& W. M. Gu, ApJS., \textbf{224}: 29(2016)
\bibitem{r2} J. J. Liu, MNRAS., \textbf{438}: 930(2014)
\bibitem{r3} J. J. Liu, \& D. M. Liu., Ap\&SS., \textbf{361}: 246(2016)
\bibitem{r4} G. M. Fuller, W. A. Fowler, and  M. J. Newman, ApJ., \textbf{42}: 447(1980)
\bibitem{r5} G. M. Fuller, W. A. Fowler, and  M. J. Newman, ApJS., \textbf{48}: 279(1982)
\bibitem{r6} M. B. Aufderheide, G. E. Brown, T. T. S. kuo, D. B. Stout, and P. Vogel., ApJ., \textbf{362}: 241(1990)
\bibitem{r7} M. B. Aufderheide, I. Fushikii,  S. E. Woosely, and D. H. Hartmanm, ApJS., \textbf{91}: 389(1994)

\bibitem{r71} M.H., Johnson,  \&  B. A., Lippmann, PhRv., \textbf{76}: 828(1949)
\bibitem{r72} W. E., Ormand, D. J., Dean, C. W. Johnson, et al., PhRvC.,\textbf{49}: 1422(1994)
\bibitem{r73} S. E., Koonin, D. J., Dean, and  K., Langanke, PhRep.,\textbf{278}: 1(1997)
\bibitem{r74} Y., Alhassid, D. J., Dean,  S. E., Koonin, et al., PhRvL.,\textbf{72}: 613(1994)

\bibitem{r8} K. Langanke \& G. Martinez-Pinedo, Phys. Lett. B., \textbf{436}: 19(1998)
\bibitem{r9} K. Langanke \& G. Martinez-Pinedo, , Nuclear Phys. A., \textbf{673}: 481(2000)
\bibitem{r10} A. Juodagalvis, K. Langanke, W. R. Hix, et al., Nuclear Phys. A., \textbf{848}: 454(2010)
\bibitem{r11} Z. F. Gao,  N. Wang, J. P. Yuan, L. Jiang, D. L. Song, ApS\&S., \textbf{332}: 129(2011)
\bibitem{r12} J. J. Liu, and Z. Q. Luo., Chin. Phys. Lett., \textbf{16}: 1861(2007)
\bibitem{r13} J. J. Liu, and Z. Q. Luo., Chin. Phys., \textbf{16}: 2671(2007)
\bibitem{r14} J. J. Liu, and Z. Q. Luo., Chin. Phys., \textbf{16}: 3624(2007)
\bibitem{r15} J. J. Liu, and Z. Q. Luo., Chin.Phys. C., \textbf{32}: 108(2008)
\bibitem{r16} J. J. Liu, and Z. Q. Luo., Comm.Theo. Phys., \textbf{49}: 239(2008)
\bibitem{r17} J. J. Liu, X. P. Kang, et al., Chin. Phys. C., \textbf{35}: 243(2011)
\bibitem{r18} J. J. Liu, Chin. Phys. C., \textbf{34}: 171(2010)
\bibitem{r19} J. J. Liu, Chin. Phys.C., \textbf{34}: 190(2010)
\bibitem{r20} J. J. Liu, RAA., \textbf{16}: 30(2016)
\bibitem{r21} J. J. Liu, Chin. Phys. C., \textbf{37}: 51018(2013)
\bibitem{r22} J. Nabi, and H. V. Klapdor-Kleingrothaus, EPJA, \textbf{337}: 339(1999)
\bibitem{r23} D. J. Dean, K. Langanke, L. Chatterjee, P. B. Radha, and M. R. Strayer, Phys. Rev. C., \textbf{58}: 536(1998)
\bibitem{r24} A. Heger,  S. E. Woosley, G. Martinez-Pinedo and K. Langanke, ApJ., 560: 307(2001)
\bibitem{r25} J. Gutierrez, E. Garcia-Berro, I. Iben, et al., ApJ., \textbf{459}: 701(1996)
\bibitem{r26} E. Bravo and D. Garcia-Senz., MNRAS., \textbf{307}: 984(1999)
\bibitem{r27} J. J. Liu, Chin. Phys. B., \textbf{19}: 099601(2010)
\bibitem{r28} N. Itoh, N. Tomizawa , M. Tamamura, et al., ApJ., \textbf{579}: 380(2002)
\bibitem{r29} Z. Q. Luo, and Q. H. Peng, ChA\&A, \textbf{25}: 1( 2001)
\bibitem{r30} J. A. Holmes, S. E. Woosley, W. A. Fowler, B. A. Zimmerman, Atomic Data and Nuclear Data Tables., \textbf{18}: 305(1986)
\bibitem{r31} F-K. Thielemann, J. W. Truran, and M. Arnould, ana..work., \textbf{525}: 540(1986)
\bibitem{r32} J. Cooperstein, and J. Wambach., Nuclear Phys. A, \textbf{420}: 591(1984)
\bibitem{r33} B. Jancovici, Nuovo Cimento, \textbf{25}: 428(1962)
\bibitem{r34} P. A. Seeger, and W. M. Howard, Nucl. Phys. A, \textbf{238}: 491(1975)


\end{thebibliography}


\end{document}